\documentclass[11pt]{article}
\usepackage{epsfig} 
\newcommand{\sect}[1]{ \section{#1} \setcounter{equation}{0} }

\newcommand{\pslash}{p \! \! \! /} 
\newcommand{\qslash}{q \! \! \! /}

\newcommand{\half}{\mbox{\small{$\frac{1}{2}$}}}

\newcommand{\MSbar}{\overline{\mbox{MS}}} 
\newcommand{\MSbars}{\overline{\mbox{\footnotesize{MS}}}}

\newcommand{\Nf}{N_{\!f}}

\begin{document}
\title{Bilinear quark operator renormalization at generalized symmetric point}
\author{J.M. Bell \& J.A. Gracey, \\ Theoretical Physics Division, \\ 
Department of Mathematical Sciences, \\ University of Liverpool, \\ P.O. Box 
147, \\ Liverpool, \\ L69 3BX, \\ United Kingdom.} 
\date{}
\maketitle 

\vspace{5cm} 
\noindent 
{\bf Abstract.} We compute Green's functions with a bilinear quark operator 
inserted at non-zero momentum for a generalized momentum configuration to two
loops. These are required to assist lattice gauge theory measurements of the
same quantity in matching to the high energy behaviour. The flavour non-singlet
operators considered are the scalar, vector and tensor currents as well as the 
second moment of the twist-$2$ Wilson operator used in deep inelastic 
scattering for the measurement of nucleon structure functions. 

\vspace{-16.5cm}
\hspace{13cm}
{\bf LTH 1077}

\newpage 

\sect{Introduction.}

The protons and neutrons of the atomic nucleus are known nowadays to be 
comprised of fundamental point-like particles called quarks. These are held in 
the nucleon bound states by the strong colour force whose quanta are gluons. 
Neither quarks nor gluons have been seen in isolation in nature due to
confinement, which is sometimes called infrared slavery. This term originates 
in part from the fact that when protons and neutrons are probed experimentally 
at high energies they behave as free fundamental entities in that asymptotic 
regime. This manifestation as apparently free particles is an apparition since 
it is not the picture at all energy scales. At low energies the colour glue 
restricts the release of isolated quarks. While high energy experiments have 
demonstrated that our theoretical understanding at a Lagrangian level is 
essentially correct, via the non-abelian gauge theory known as Quantum 
Chromodyanmics (QCD), how protons and neutrons actually form and become the 
observed degrees of freedom is still not fully resolved. One way of studying 
properties of the nucleon substructure is to model the internal quark behaviour
numerically. This goes well beyond the perturbative region of QCD and requires 
the technique known as lattice gauge theory. The proton and neutron structure 
functions can be constructed from models of the non-perturbative behaviour with
lattice gauge theory input data and in the main this has progressed our 
understanding of nucleon structure.

Given this overview there is a technical side to actually extracting the
information required for the nucleon structure functions. Based on the operator
product expansion one computes moments of the structure functions which are
related to Green's functions involving twist-$2$ operators, \cite{1}. For the 
flavour non-singlet sector of the structure functions the $n$th moment of these
operators are quark anti-quark operators involving $(n-1)$ covariant 
derivatives. Their renormalization properties have been known for many years 
with the provision of the one, two and three loop $\MSbar$ anomalous 
dimensions, \cite{1,2,3}. However, the dependence of the Green's function 
itself on momenta and the strong coupling constant is what is necessary for 
proton and neutron structure analysis. As noted this is the quantity computed 
on the lattice non-perturbatively. However, these calculations are intensely 
numerical and the measurement of a central value at a particular momentum is 
only part of the work. The second aspect is to refine the error bounds so that 
they are small. One aspect of that analysis is that in addition the behaviour 
of the measured Green's function should match onto the high energy piece. The 
latter is determined in perturbation theory, \cite{1,2,3}. Indeed this has been
the modus operandi for the last decade or so. Multiloop evaluation of the same 
Green's function can be determined to two and three loops. For the most part 
the latter order is for the momentum configuration where the operator is at 
zero momentum insertion in a quark $2$-point function. However, more recently 
the focus has switched to forward matrix elements which requires a non-zero 
momentum flowing through the inserted operator. This relates to generalized 
parton distribution functions.

In this instance the supporting perturbative computations have been carried out
at the fully symmetric momentum configuration but only to two loops. Such
results, \cite{4,5,6,7,8} have proved important in lattice analyses. For 
example, see 
\cite{9,10,11,12,13,14,15,16,17,18,19,20,21,22,23,24,25,26,27,28,29,30} for a 
non-exhaustive representation. While this has provided a useful start for such 
forward matrix element analyses, it is slightly restrictive. In computing at 
the fully symmetric point the squares of all incoming external momenta are 
equal, \cite{4,5,7,8}. That value for the operator insertion is no different 
from that of the external quark legs on the Green's function. Instead a more 
general configuration would be preferable where the momenta flowing into the 
operator itself is not tied to those of the other legs. Therefore, it is the 
purpose of this article to address this problem. We will compute the Green's 
function of the relevant quark operators, which are twist-$2$ flavour 
non-singlet operators. The particular momentum configuration considered will be
called the interpolating momentum (IMOM) subtraction point. This is in 
contradistinction to the symmetric momentum (SMOM) subtraction point of 
\cite{4,5,7,8}. The origin of our IMOM nomenclature is due to the appearance of
a parameter, $\omega$, which will be free. It parametrizes the momentum 
transfer at the operator insertion. Our results will all be functions of 
$\omega$ and will provide a freedom for the lattice to tune their analysis. For
instance, it may transpire that for certain values of $\omega$ the error bounds
are tighter than for others. Our choice of IMOM configuration, which is a 
non-exceptional one, is not original in that it was considered in \cite{6} for 
the renormalization of the quark mass operator. As a first stage we will 
reconstruct that computation \cite{6} partly as a check on our computational 
setup but also because we will give the full decomposition of the Green's 
function into its Lorentz basis. We repeat the process for both the vector 
current and tensor quark bilinear. The former is considered because of its 
later relation to the second moment of the twist-$2$ flavour non-singlet 
operator of deep inelastic scattering. The tensor operator is evaluated because
of its relation to the vector meson decay constant, \cite{31}, for instance. 
Our analysis is completed by the calculation which is that of the full set of 
second moment twist-$2$ operators. With the increase in Lorentz indices in the 
case of this set of operators, the basis of tensors for the Green's function 
itself enlarges. However, we will compute the full result. This is necessary as
lattice regularized QCD makes measurements in different directions in order to 
decipher the signal which contributes to the matrix element for the structure 
function. Throughout we work in the $\MSbar$ scheme and do not consider any of 
the hybrid renormalization schemes which were introduced in earlier work such 
as the modified regularization invariant (RI${}^\prime$), \cite{32,33}, or 
momentum subtraction (MOM) schemes, \cite{34,35}. This is partly as lattice 
motivated schemes can readily be converted to the reference $\MSbar$ scheme but
also because there is a large degree of freedom in defining a non-minimal 
momentum subtraction scheme which derives from RI${}^\prime$.

The paper is organized as follows. The definition of the operators we consider
and the quantum field theoretic formalism used is introduced in section $2$. 
This includes a discussion on the technical details of the computation. The 
subsequent two sections contain the main results with section $3$ concentrating
on the basic bilinear quark operators while section $4$ focuses on the 
twist-$2$ Wilson operator second moment. We provide concluding remarks in 
section $5$. An appendix records the tensor basis for the Green's function of 
each operator together with the explicit projection matrices which are 
$\omega$-dependent. The latter are required to extract the coefficient of each 
basis tensor as a function of $\omega$.  

\sect{Background.}

We begin by briefly reviewing the formalism we use which will be based on
earlier articles, \cite{5,6,7,8}. We focus on the main differences due to the
more general external momentum configuration. We refer the interested reader to
these articles for more detail but our notation will be the same as \cite{7,8}.
First, for shorthand and for labelling purposes the operators we consider are 
\begin{equation}
S ~ \equiv ~ \bar{\psi} \psi ~~~,~~~
V ~ \equiv ~ \bar{\psi} \gamma^\mu \psi ~~~,~~~
T ~ \equiv ~ \bar{\psi} \sigma^{\mu\nu} \psi ~~~,~~~
W_2 ~ \equiv ~ {\cal S} \bar{\psi} \gamma^\mu D^\nu \psi ~~~,~~~
\partial W_2 ~ \equiv ~ {\cal S} \partial^\mu \left( \bar{\psi} \gamma^{\nu}
\psi \right) 
\label{oplabel}
\end{equation}
where $\sigma^{\mu\nu}$~$=$~$\half [\gamma^\mu , \gamma^\nu]$. The letters $S$,
$V$, $T$, $W_2$ and $\partial W_2$ denote scalar, vector, tensor, twist-$2$ 
flavour non-singlet second moment Wilson operator and total derivative of the 
latter respectively. In the case of $W_2$ and $\partial W_2$ each operator is
symmetrized with respect to the Lorentz indices and is also traceless in
$d$-dimensions. The operator for this latter procedure is denoted by ${\cal S}$
and if  
\begin{equation}
{\cal O}^{W_2}_{\mu\nu} ~=~ \bar{\psi} \gamma_\mu D_\nu \psi 
\end{equation}
for instance, then, \cite{1,2}, 
\begin{equation}
{\cal S} {\cal O}^{W_2}_{\mu\nu} ~=~ {\cal O}^{W_2}_{\mu\nu} ~+~
{\cal O}^{W_2}_{\nu\mu} ~-~ \frac{2}{d} \eta_{\mu\nu}
{\cal O}^{W_2\,\sigma}_{\sigma} 
\end{equation}
in $d$-dimensions. For each of the operators in (\ref{oplabel}) we will 
evaluate the Green's function
\begin{equation}
\left. \left\langle \psi(p) {\cal O}^L_{\mu_1 \ldots \mu_{n_L}}(r) 
\bar{\psi}(q) \right\rangle \right|_\omega
\label{gfun}
\end{equation}
where $L$~$=$~$S$, $V$, $T$, $W_2$ or $\partial W_2$ and
\begin{equation}
p ~+~ q ~+~ r ~=~ 0
\end{equation}
by conservation of momentum. The restriction here is shorthand notation for the
evaluation of the squares of the three external momentum at the interpolating
momentum subtraction point we are interested in. In particular the squared
momentum flowing into the operator will always be the one which is distinct 
from the quark leg momenta given that we take, \cite{6}, 
\begin{equation}
p^2 ~=~ q^2 ~=~ -~ \mu^2 ~~,~~ r^2 ~=~ -~ \omega \mu^2
\label{imom1}
\end{equation}
which imply
\begin{equation}
pq ~=~ \left[ 1 - \frac{\omega}{2} \right] \mu^2 ~~,~~
pr ~=~ qr ~=~ \frac{\omega}{2} \mu^2 ~.
\label{imom2}
\end{equation}
The $\omega$~$\rightarrow$~$1$ limit will correspond to the symmetric point
used in SMOM. On a note of caution the $\omega$~$\rightarrow$~$0$ limit will be
infrared unsafe and will not be considered in any analysis. Equally the value
of $\omega$~$=$~$4$ corresponds to a collinear singularity. Therefore we will
only allow $\omega$ to be in the range $0$~$<$~$\omega$~$<$~$4$. The number of 
free Lorentz indices $n_L$ for each operator is indicated in (\ref{oplabel}). 

{\begin{table}[ht]
\begin{center}
\begin{tabular}{|c||c|c|c|c|}
\hline
$L$ & $S$ & $V$ & $T$ & $W_2$ \\
\hline
$N_L$ & $2$ & $6$ & $8$ & $10$ \\
\hline
\end{tabular}
\end{center}
%\vspace{0.3cm}
\begin{center}
{Table 1. Number of Lorentz basis elements, $N_L$, for each operator insertion 
$L$.}
\end{center}
\end{table}}
 
In order to provide the structure of the Green's function which lattice gauge 
theory requires for the measurements and matching we decompose (\ref{gfun})
into its Lorentz basis at the IMOM subtraction point which we will sometimes
refer to as an asymmetric point. Specifically we write, \cite{7,8}, 
\begin{equation}
\left. \left\langle \psi(p) {\cal O}^L_{\mu_1 \ldots \mu_{n_L}}(r)
\bar{\psi}(q) \right\rangle \right|_\omega ~=~ \sum_{k=1}^{N_L}
{\cal P}^L_{(k) \, \mu_1 \ldots \mu_{n_L} }(p,q) \,
\left. \Sigma^L_{(k)}(p,q) \right|_\omega
\end{equation}
where $\left. \Sigma^L_{(k)}(p,q) \right|_\omega$ are the scalar amplitudes at
the IMOM point associated with each respective Lorentz tensor basis element, 
${\cal P}^L_{(k) \, \mu_1 \ldots \mu_{n_L} }(p,q)$. The number of tensors in
the basis, $N_L$, is given in Table $1$ and the basis for each of the operators 
considered here is given explicitly in Appendix A. To determine the scalar 
amplitudes we extend the projection method used in previous articles to the 
specific momentum configuration of interest. In particular the $k$th amplitude 
can be extracted by applying the projection matrix ${\cal M}^L_{kl}$ to the 
Green's function (\ref{gfun}) itself. In other words 
\begin{equation}
\left. \Sigma^L_{(k)}(p,q) \right|_\omega ~=~ {\cal M}^L_{kl}
{\cal P}^{L ~\, \mu_1 \ldots \mu_{n_L}}_{(l)}(p,q) \left. \left(
\left\langle \psi(p) {\cal O}^L_{\mu_1 \ldots \mu_{n_L}}(-p-q) \bar{\psi}(q)
\right\rangle \right) \right|_\omega
\end{equation}
where there is a sum over the amplitude label $l$. The projection matrix is 
deduced from the basis tensors and is defined as the inverse of the matrix
given by, \cite{7,8},
\begin{equation}
{\cal N}^L_{kl} ~=~ \left. {\cal P}^L_{(k) \, \mu_1 \ldots \mu_{n_L}}(p,q)
{\cal P}^{L ~\, \mu_1 \ldots \mu_{n_L}}_{(l)}(p,q) \right|_\omega ~.
\end{equation}
The entries in ${\cal M}^L_{kl}$ and ${\cal N}^L_{kl}$ are polynomials in $d$
and recorded in the former case for each operator in Appendix A. Due to the
asymmetry of the momentum configuration both matrices are $\omega$ dependent. 
As is evident several elements are singular at $\omega$~$=$~$0$ reflecting the 
infrared issue at a nullified operator insertion. 

Having discussed the quantum field theory formalism underlying the problem we
now summarize the practicalities behind determining the amplitudes. We have
evaluated all the one and two loop Feynman integrals using an automatic
symbolic manipulation programme written in the language {\sc Form}, 
\cite{36,37}. This is the most efficient tool to handle the large amounts of 
intermediate algebra. The algorithm is initiated from the electronic 
representation of the Feynman graphs generated with the {\sc Qgraf} package, 
\cite{38}. These graphs are adapted by automatically labelling with colour, 
spinor, Lorentz and flavour indices in a {\sc Form} module prior to applying 
the projection matrix to isolate the individual scalar amplitudes within 
several other {\sc Form} modules. The bulk of the evaluation centres on the 
integration of the large set of Feynman integrals contributing to a graph after
scalar products of internal and external momenta are written in terms of the 
propagators, respecting the relations (\ref{imom1}) and (\ref{imom2}). In 
arranging the scalar products in this way some integrals are produced with 
irreducible numerators. By this we mean that there was either no propagator of 
that form in the original topology or a propagator has a negative power. To 
proceed requires using integration by parts enshrined in the Laporta algorithm,
\cite{39}. This is a technique which systematically constructs algebraic 
relations between scalar integrals with or without irreducible propagators. The
resulting tower of relations, while over-redundant, can then be solved in terms
of a minimal number of basic scalar integrals. Termed masters their 
$\epsilon$-expansion can only be deduced by non-integration by parts methods. 
All our calculations are carried out using dimensional regularization in 
$d$~$=$~$4$~$-$~$2\epsilon$ dimensions. However, for the asymmetric momentum
configuration we are considering it transpires that the various master
integrals are known to the required order in $\epsilon$ from various sources,
\cite{40,41,42,43,6}. The various polylogarithm functions which appear in the 
finite part of (\ref{gfun}) are noted in the next section. For our automatic 
evaluation we have generated a database of integrals using the {\sc Reduze}, 
\cite{44}, implementation of the Laporta algorithm. The output is readily 
converted into {\sc Form} syntax and thence into a {\sc Form} module. Once each 
contributing graph has passed through the integration algorithm they are summed
and the Green's function rendered finite. This is carried out automatically
using the process provided in \cite{45}. We work with a bare coupling constant 
and gauge parameter, $\alpha$, in the evaluation of all Feynman diagrams and 
their renormalized counterparts are introduced by the usual rescaling. This 
introduces the counterterms without having to perform subtractions which would 
be difficult to implement in a symbolic manipulation programme. There are 
various checks on the final expression for the Green's function for each 
operator. First, we use an arbitrary linear covariant gauge. Therefore, as each
operator is gauge invariant its $\MSbar$ renormalization constant will be 
independent of the gauge fixing parameter $\alpha$. Moreover, it cannot depend 
on $\omega$ and we note that each operator passed this test as well as agreeing
with the known two loop renormalization constants. The other main check was 
that we recovered the known results, \cite{4,5,6,7,8} for each Green's function
in the $\omega$~$\to$~$1$ limit and therefore we are confident that our results
are not incorrect. 

\sect{Basic operators.}

We now turn to the discussion of our results. As a check on our {\sc Form} code
we have reproduced the asymmetric point results of \cite{6} for the scalar 
current $S$. In particular we found agreement with the conversion function for
the anomalous dimension of the mass operator from the $\MSbar$ scheme to the 
scheme defined in \cite{6}. This scheme was termed the IMOM scheme, \cite{6},
to distinguish it from the MOM or SMOM schemes. We will use the same 
nomenclature here. However, as we used a different point of view for the 
renormalization of $\alpha$ in the RI${}^\prime$ scheme aspect of the IMOM 
scheme the results will only tally in the Landau gauge which is the only gauge 
of interest for lattice matching anyway. In the attached data file we record 
the second channel amplitude in addition to that for channel $1$ for 
completeness. One of the main results of \cite{6} was the conversion function. 
Although this was presented in \cite{6} we record the same quantity here in 
order to compare the different gauge parameter dependence. We found
\begin{eqnarray}
C_{\bar{\psi}\psi}(a,\alpha) &=&
1 + \left[ \Phi_{(1)\,\omega,\omega} \alpha + 3 \Phi_{(1)\,\omega,\omega} 
- 2 \alpha - 8 \right] \frac{C_F a}{2} \nonumber \\ 
&& +~ \left[ -~ 36 \ln(\omega) \Phi_{(1)\,\omega,\omega} \alpha \omega C_F
- 12 \ln(\omega) \Phi_{(1)\,\omega,\omega} \omega C_A
- 60 \ln(\omega) \Phi_{(1)\,\omega,\omega} \omega C_F
\right. \nonumber \\
&& \left. ~~~~
+ 12 \Omega_{(2)\,\omega,\omega} \omega C_A
- 24 \Omega_{(2)\,\omega,\omega} \omega C_F
- 24 \Omega_{(2)\,1,\omega} \omega C_A
+ 48 \Omega_{(2)\,1,\omega} \omega C_F
\right. \nonumber \\
&& \left. ~~~~
+ 6 \Phi_{(1)\,\omega,\omega}^2 \alpha^2 \omega C_F
+ 36 \Phi_{(1)\,\omega,\omega}^2 \alpha \omega C_F
+ 12 \Phi_{(1)\,\omega,\omega}^2 \omega C_A
- 24 \Phi_{(1)\,\omega,\omega}^2 C_A 
\right. \nonumber \\
&& \left. ~~~~
+ 30 \Phi_{(1)\,\omega,\omega}^2 \omega C_F
+ 48 \Phi_{(1)\,\omega,\omega}^2 C_F
+ 9 \Phi_{(1)\,\omega,\omega} \alpha^2 \omega C_A
- 24 \Phi_{(1)\,\omega,\omega} \alpha^2 \omega C_F
\right. \nonumber \\
&& \left. ~~~~
+ 42 \Phi_{(1)\,\omega,\omega} \alpha \omega C_A
- 48 \Phi_{(1)\,\omega,\omega} \alpha \omega C_F
+ 385 \Phi_{(1)\,\omega,\omega} \omega C_A
- 168 \Phi_{(1)\,\omega,\omega} \omega C_F
\right. \nonumber \\
&& \left. ~~~~
- 80 \Phi_{(1)\,\omega,\omega} \omega T_F \Nf
- 24 \Phi_{(2)\,\omega,\omega} \alpha \omega C_F
- 24 \Phi_{(2)\,\omega,\omega} \omega C_A
- 24 \Phi_{(2)\,\omega,\omega} \omega C_F
\right. \nonumber \\
&& \left. ~~~~
+ 24 \Phi_{(2)\,1,\omega} \omega^2 C_A
- 18 \alpha^2 \omega C_A
+ 24 \alpha^2 \omega C_F
- 84 \alpha \omega C_A
+ 96 \alpha \omega C_F
\right. \nonumber \\
&& \left. ~~~~
+ 288 \omega \zeta_3 C_A
- 1285 \omega C_A
+ 57 \omega C_F
+ 332 \omega T_F \Nf \right] \frac{C_F a^2}{24 \omega} ~+~ O(a^3) 
\end{eqnarray}
where $a$ and $\alpha$ are $\MSbar$ variables and $a$~$=$~$g^2/(16\pi^2)$ in
terms of the gauge coupling constant $g$. Also $C_F$, $C_A$ and $T_F$ are the 
usual group theoretic quantities for a Lie group and $\zeta_z$ is the Riemann 
zeta function. Our shorthand notation for various functions is
\begin{eqnarray}
\Phi_{(n)\,1,\omega} &=& \Phi_{(n)} \left(1,\omega\right) ~~~,~~~
\Phi_{(n)\,\omega,\omega} ~=~ 
\Phi_{(n)} \left( \frac{1}{\omega},\frac{1}{\omega} \right) \nonumber \\
\Omega_{(n)\,1,\omega} &=& \Omega_{(n)} \left(1,\omega\right) ~~~,~~~
\Omega_{(n)\,\omega,\omega} ~=~ 
\Omega_{(n)} \left( \frac{1}{\omega},\frac{1}{\omega} \right) ~. 
\end{eqnarray}
These arise from the various underlying master integrals and were evaluated
explicitly in terms of polylogarithm functions in \cite{40,41,42,43}. As we 
will be representing the results for the other operators of interest in 
graphical form within the article, for completeness we have plotted the channel
$1$ scalar operator amplitude in Figure $1$. There we show the one and two loop
corrections as a function of $\omega$ for values of $\Nf$ with 
$3$~$\leq$~$\Nf$~$\leq$~$6$ at a specific value of $\alpha_s$. This is for
illustration as the full results are available for the interested reader in the
data file. In Figure $1$ and subsequent figures we note that we use the
notation $L_{mnl}$ in the legend to denote the amplitude of the operator $L$. 
The subscript $mnl$ labels the Lorentz channel number, the number of quark
flavours and the loop order respectively.

{\begin{figure}
\includegraphics[width=7.6cm,height=7cm]{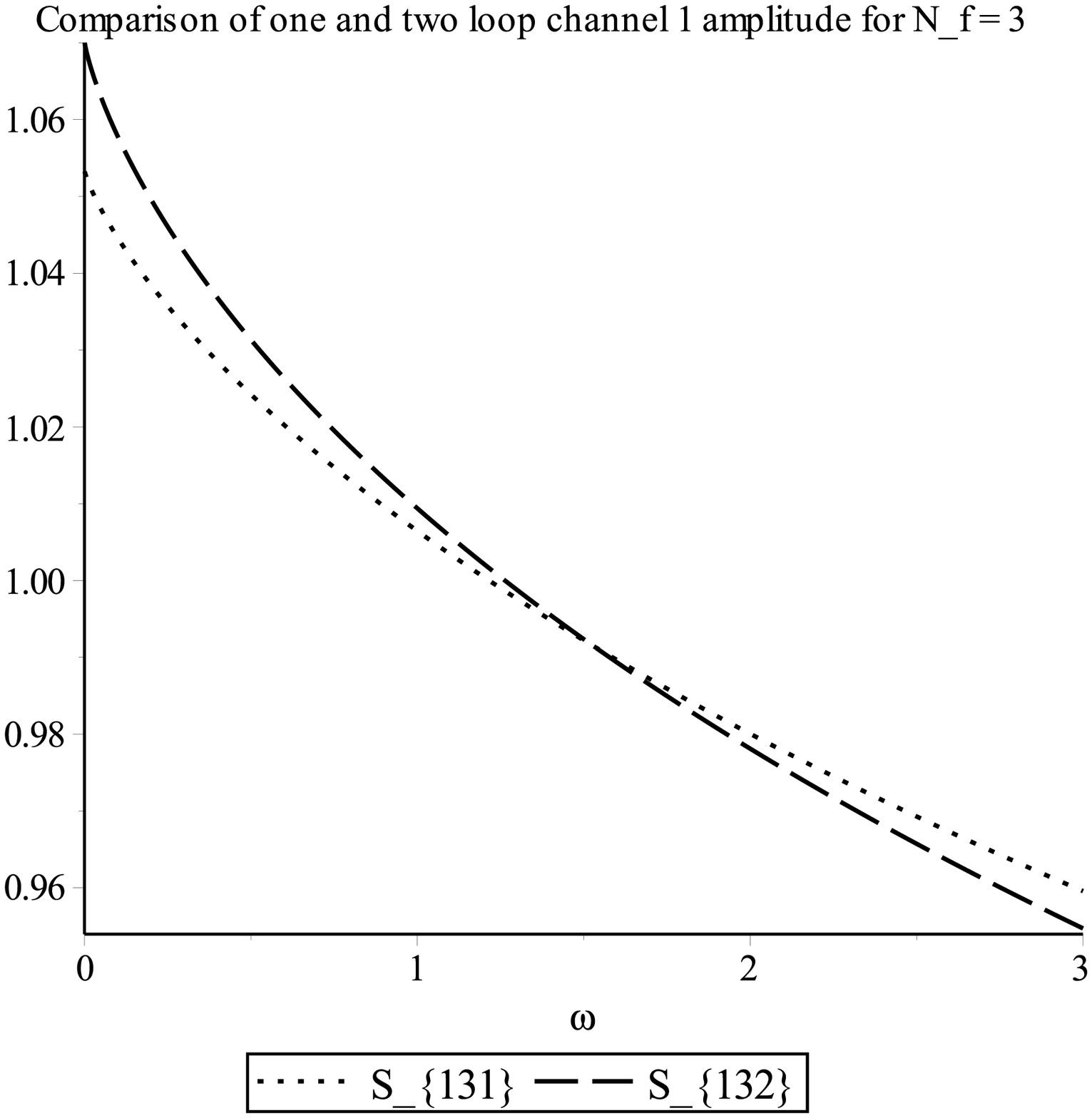}
~~~~
\includegraphics[width=7.6cm,height=7cm]{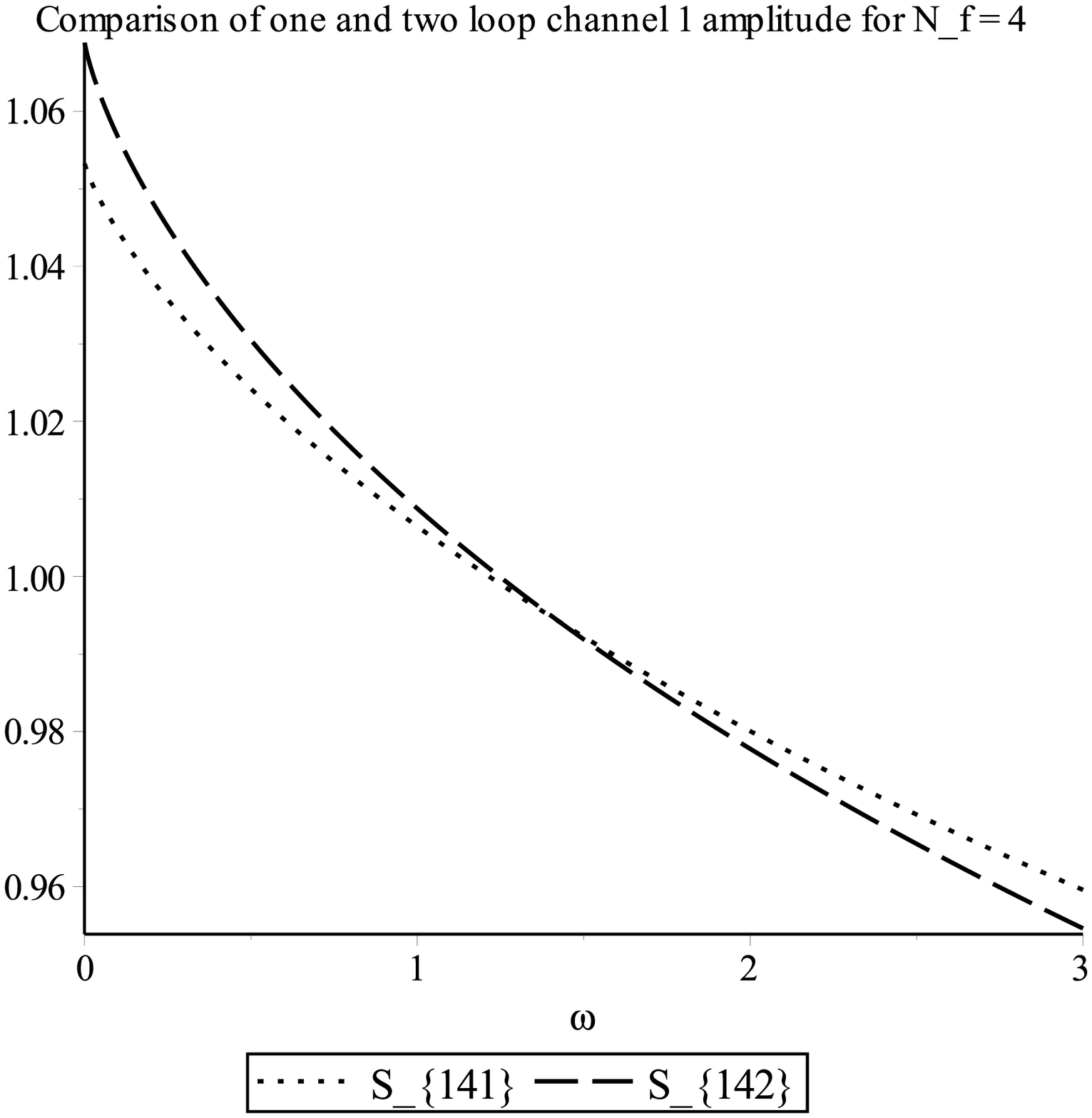}
\vspace{0.8cm}
\includegraphics[width=7.6cm,height=7cm]{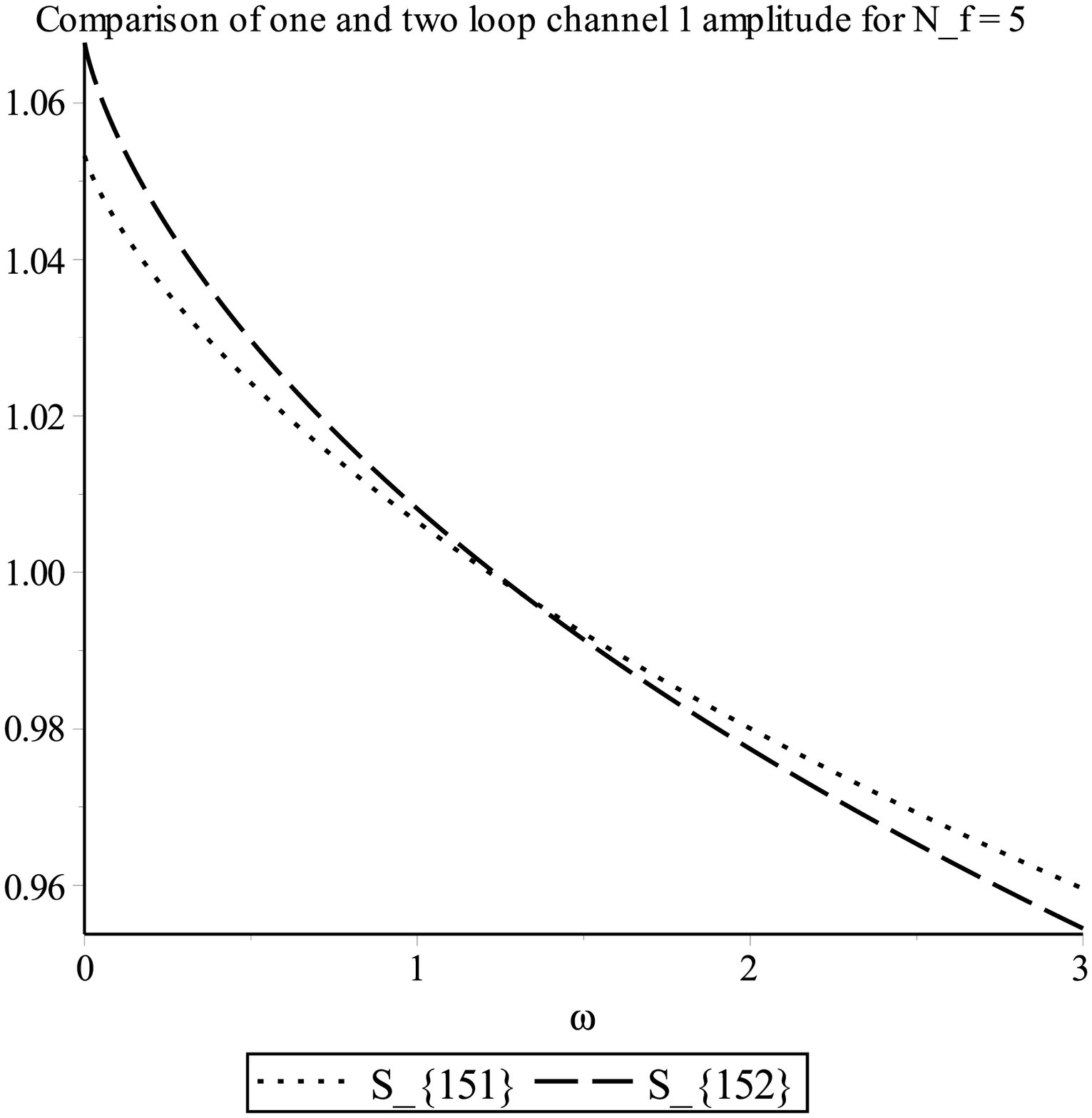}
\quad
\includegraphics[width=7.6cm,height=7cm]{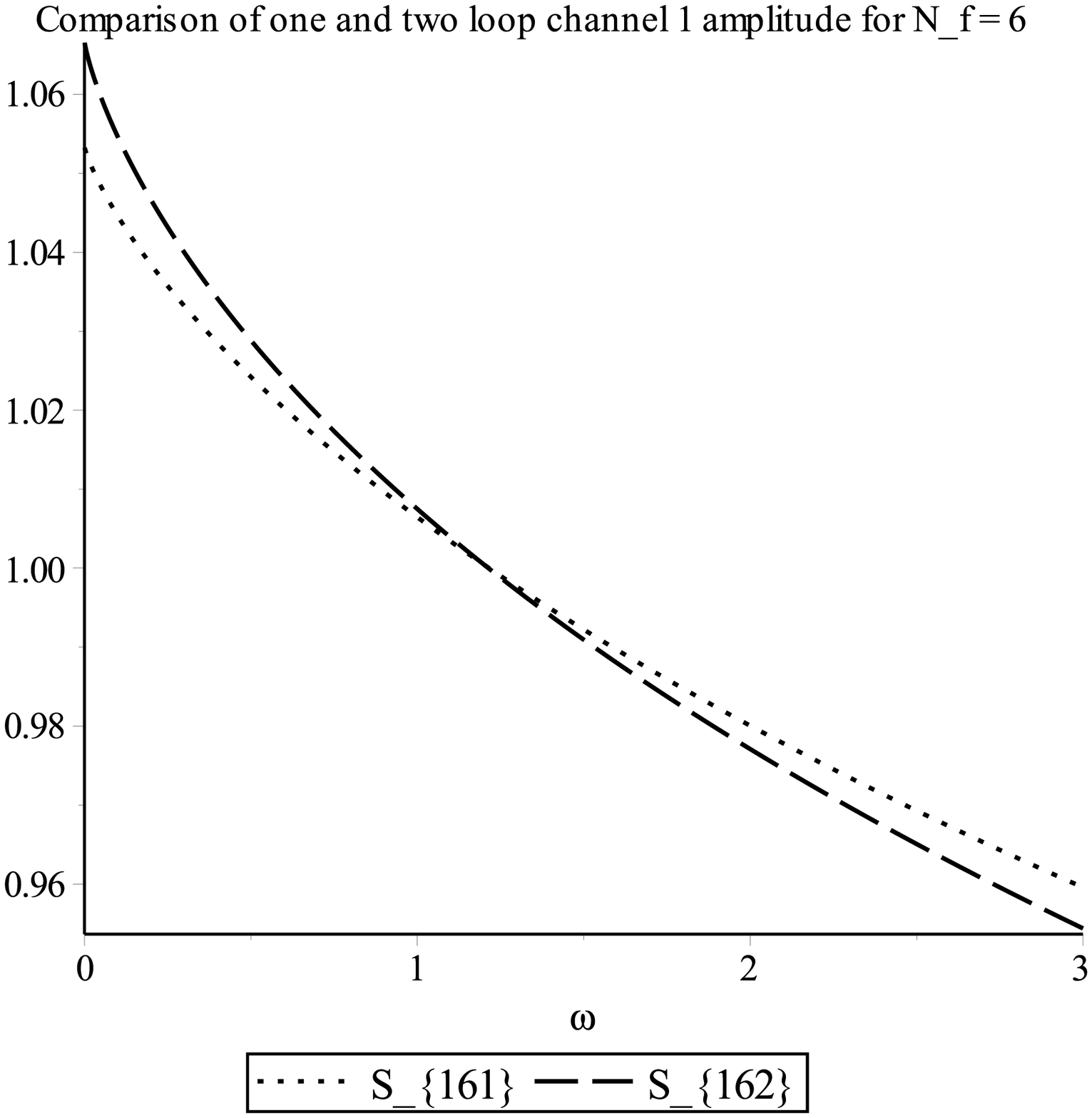}
\caption{Channel $1$ amplitude for scalar operator for different $\Nf$ values
and $\alpha_s$~$=$~$0.125$.}
\end{figure}}

The renormalization of the vector operator requires care due to its underlying
significance as a conserved physical current. In other words due to charge 
conservation in the quantum theory 
$\partial_\mu \left( \bar{\psi} \gamma^\mu \psi \right)$~$=$~$0$. So to extract 
the correct renormalization constant for $V$ one has to express this condition,
which equates effectively to the Slavnov-Taylor identity, in the context of the
Green's function with the operator inserted at non-zero momentum. A 
decomposition into the full Lorentz tensor basis is necessary for this. Once
the combination of amplitudes is constructed which corresponds to the
divergence free current condition then the vector current renormalization
constant can be set. Specifically, we require the combination
\begin{equation}
\left. \Sigma^{V}_{(1)}(p,q) \right|_{\MSbars} ~-~
\frac{1}{2} \left. \Sigma^{V}_{(2)}(p,q) \right|_{\MSbars} ~-~ \frac{1}{2}
\left. \Sigma^{V}_{(5)}(p,q) \right|_{\MSbars} 
\end{equation}
to be finite in the $\MSbar$ scheme. This is in contrast to the renormalization
of the other operators we consider here. In those cases we ensure that the
channel in the Lorentz decomposition of the Green's function containing the
tree is finite in the $\MSbar$ scheme. However, since the vector current is a 
physical operator then its renormalization is trivial in {\em all}
renormalization schemes. In other words $Z^V$~$=$~$1$ and consequently 
$\gamma^V(a)$~$=$~$0$ to all orders. For our interpolating subtraction point 
computation we have checked that this is indeed the case to two loops which 
plays an important check on the calculation. However, we have summarized the 
renormalization process for this specific operator to ensure that if others 
choose to renormalize in a scheme other than $\MSbar$ then the Slavnov-Taylor 
identity is respected. Another check on the results is that the general
relations, \cite{7}, 
\begin{equation}
\left. \Sigma^{V}_{(2)}(p,q) \right|_{\omega = 1} ~=~ 
\left. \Sigma^{V}_{(5)}(p,q) \right|_{\omega = 1} ~~~,~~~
\left. \Sigma^{V}_{(3)}(p,q) \right|_{\omega = 1} ~=~ 
\left. \Sigma^{V}_{(4)}(p,q) \right|_{\omega = 1} 
\label{Vrel}
\end{equation}
have to be satisfied in the $\MSbar$ scheme. These relations follow from the 
fact that the Green's function is symmetric under the interchange of $p$ and 
$q$ in the external legs. Their forms were not assumed at the outset but 
emerged naturally within the computation and acts as a useful check. Given that
our amplitudes fulfil the general criterion we have checked that in the 
$\omega$~$\to$~$1$ limit the fully symmetric point results correctly emerge. 
With these considerations we have plotted the channel $1$ amplitudes in Figure 
$2$ for $\alpha_s$~$=$~$0.125$. It is evident that for the value of $\alpha_s$ 
the variation from one to two loops across the ranges of $\omega$ and $\Nf$ is 
not large. In fact for these ranges it is no more than $0.5\%$. This is not 
unexpected since for this particular value of the coupling constant
perturbation theory is expected to be a solid approximation. 

{\begin{figure}
\includegraphics[width=7.6cm,height=7cm]{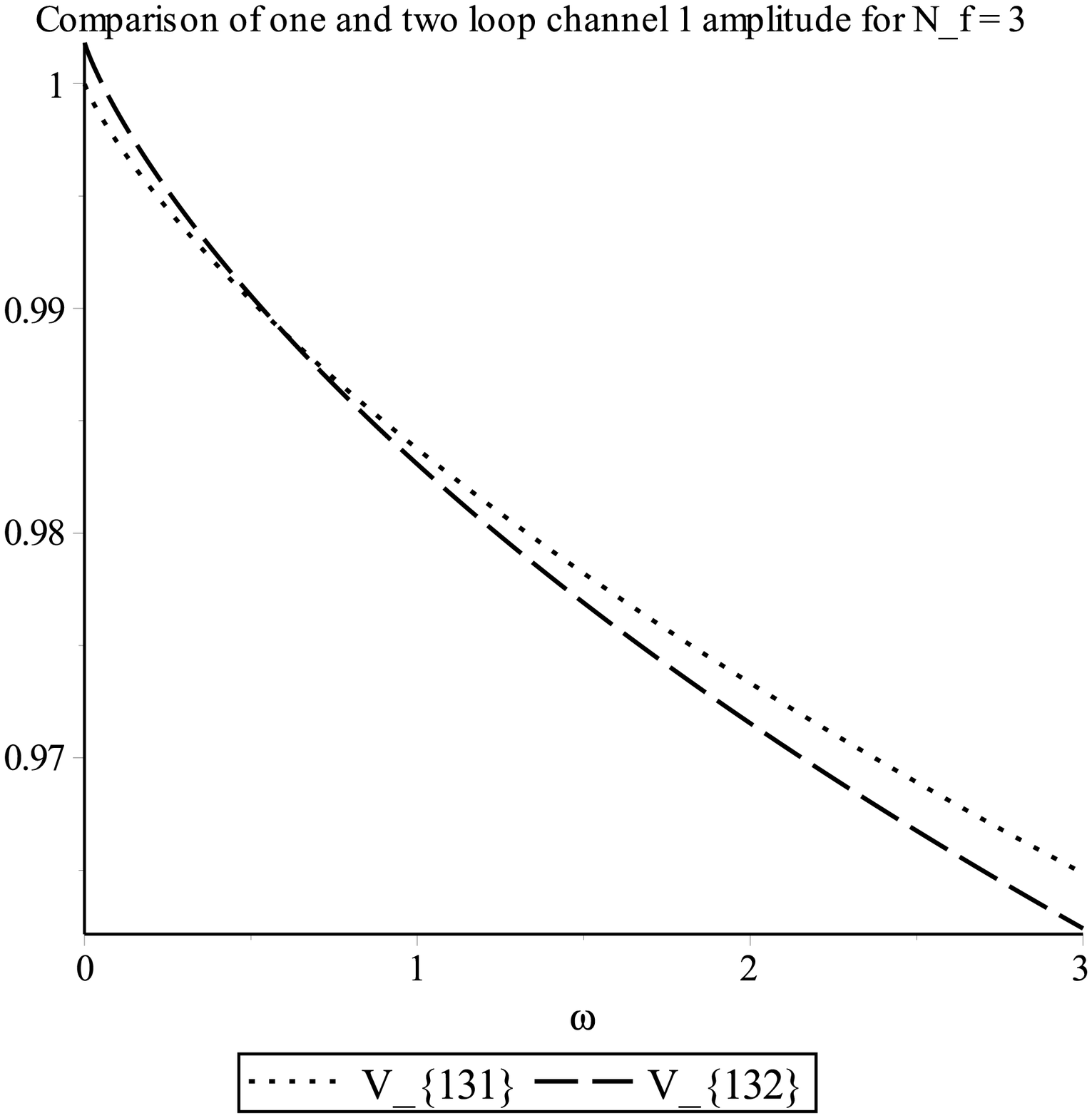}
~~~~
\includegraphics[width=7.6cm,height=7cm]{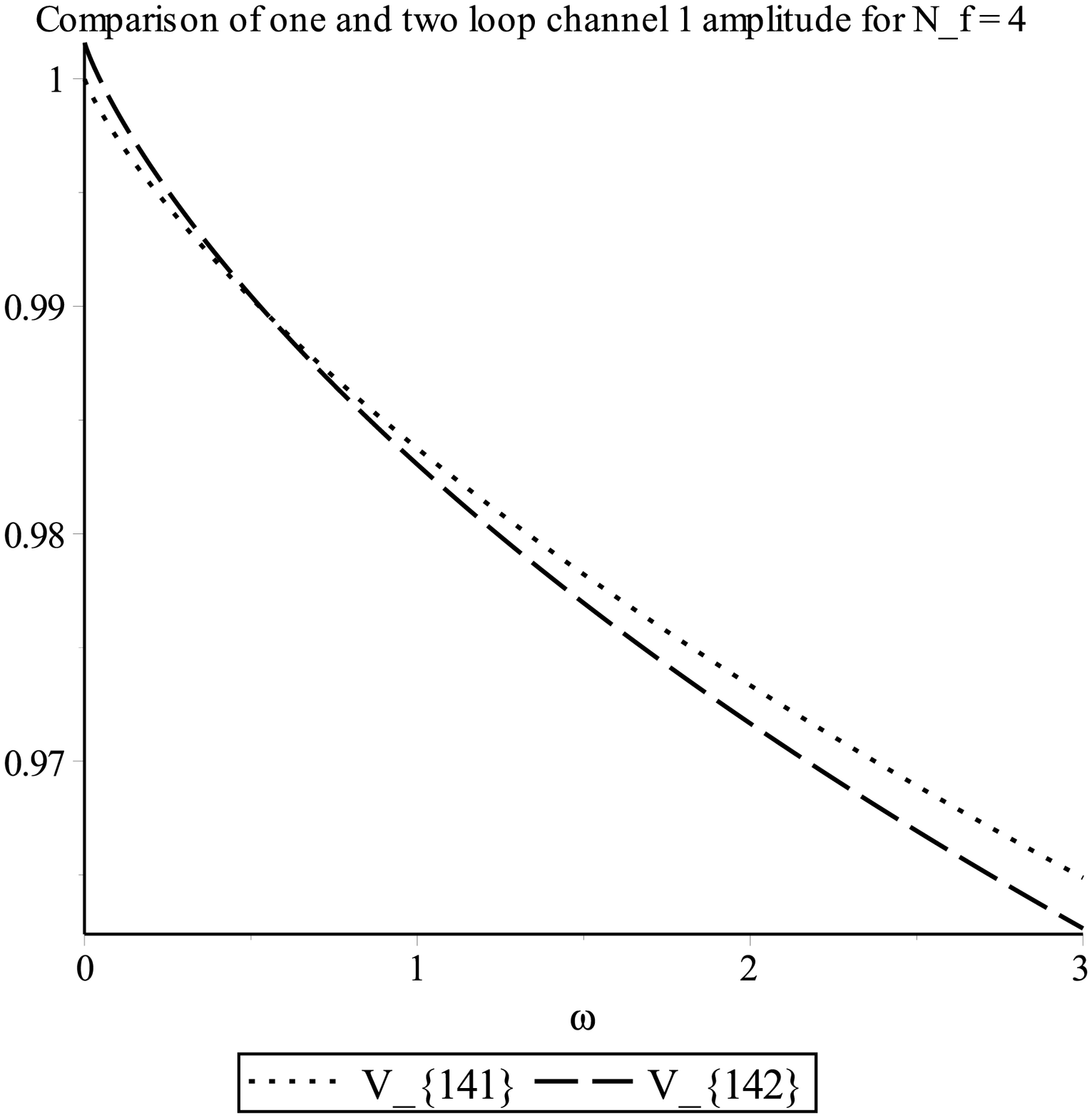}
\vspace{0.8cm}
\includegraphics[width=7.6cm,height=7cm]{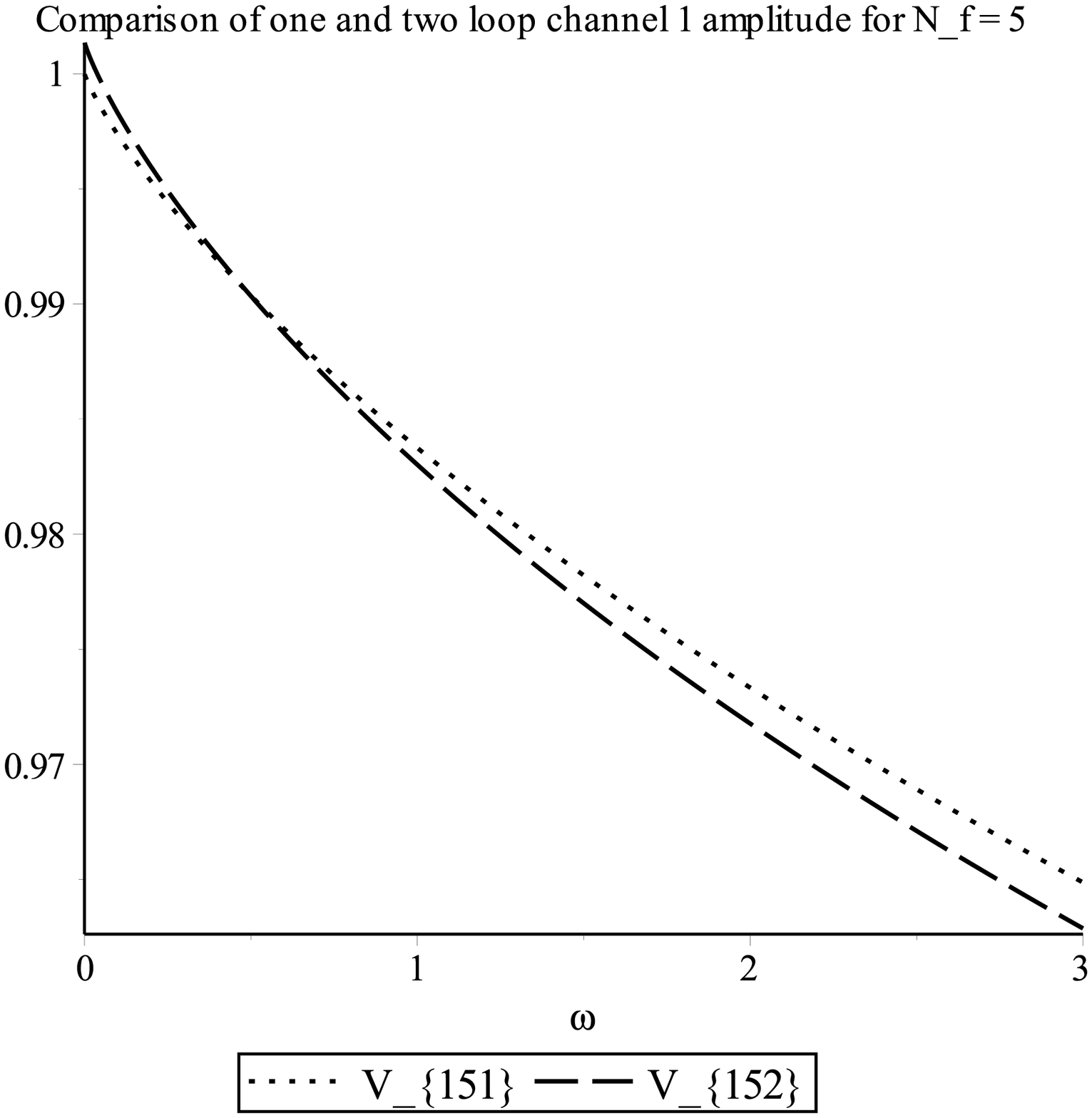}
\quad
\includegraphics[width=7.6cm,height=7cm]{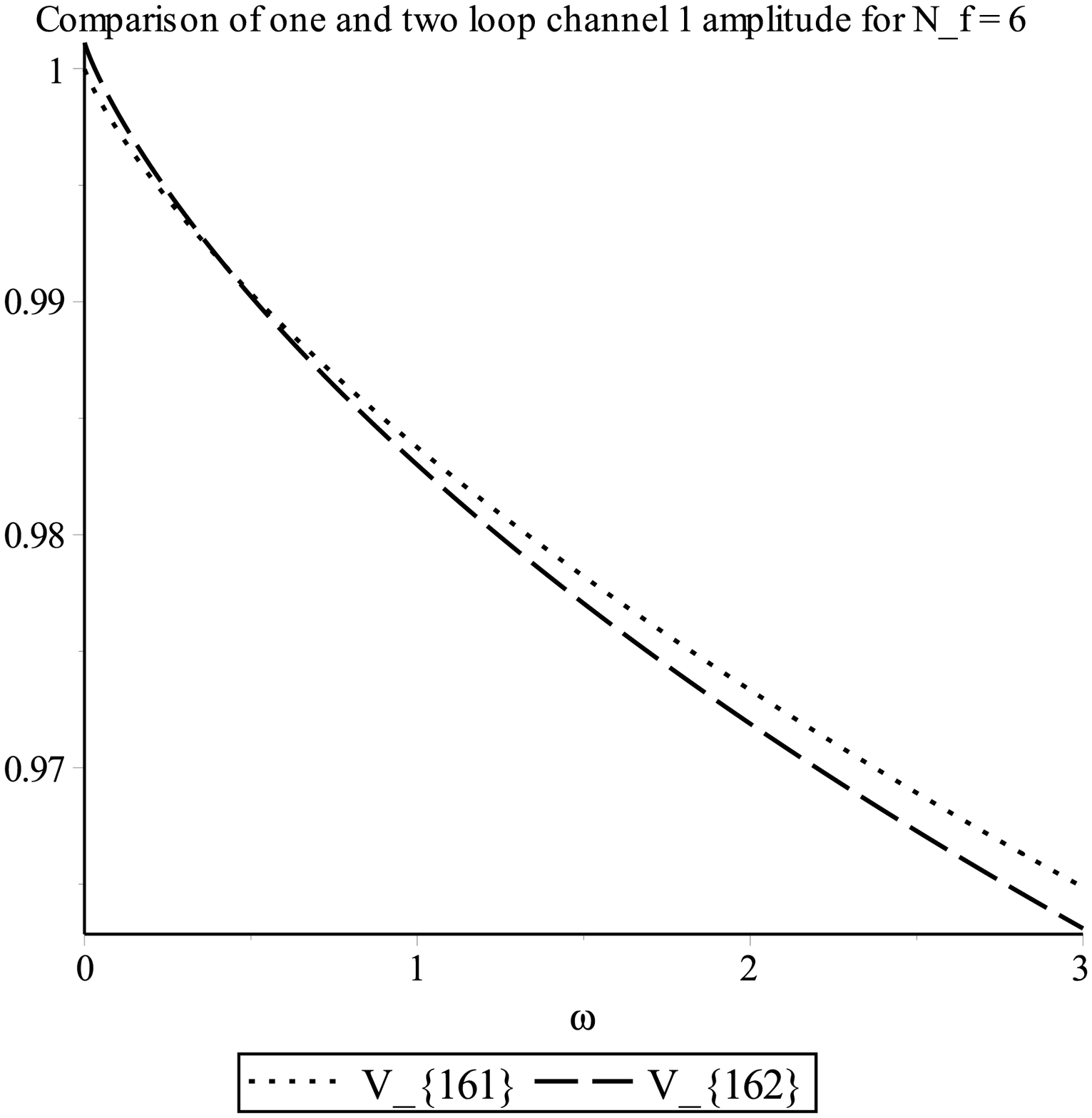}
\caption{Channel $1$ amplitude for vector current for different $\Nf$ values
and $\alpha_s$~$=$~$0.125$.}
\end{figure}}

Finally, we complete this section by considering the tensor current. For its 
renormalization we have checked that the correct $\omega$-independent two loop 
$\MSbar$ renormalization constant, \cite{46}, first emerges at the 
$\omega$-dependent subtraction point. The actual divergence in $\epsilon$ being
present only in channel $1$ as expected from renormalizability. The emergence
of the known $\MSbar$ renormalization constant is a useful check on the 
{\sc Form} setup. Concerning the other channels a similar left-right symmetry 
in the original Green's function to the vector case is present corresponding 
to, \cite{7}, 
\begin{equation}
\left. \Sigma^{T}_{(3)}(p,q) \right|_{\omega = 1} ~=~ 
\left. \Sigma^{T}_{(6)}(p,q) \right|_{\omega = 1} ~~~,~~~
\left. \Sigma^{T}_{(4)}(p,q) \right|_{\omega = 1} ~=~ 
\left. \Sigma^{T}_{(5)}(p,q) \right|_{\omega = 1} 
\end{equation}
in general. We have checked that these amplitudes satisfy the relations at two 
loops in the $\MSbar$ scheme. An illustration of the behaviour of the channel 
$1$ amplitude is given in Figure $3$. While the form of the plots appear
different from the previous two cases it should be noted that the vertical
scale is more finely grained in the tensor case. So the behaviour close to
$\omega$~$=$~$0$ is amplified.
 
{\begin{figure}
\includegraphics[width=7.6cm,height=7cm]{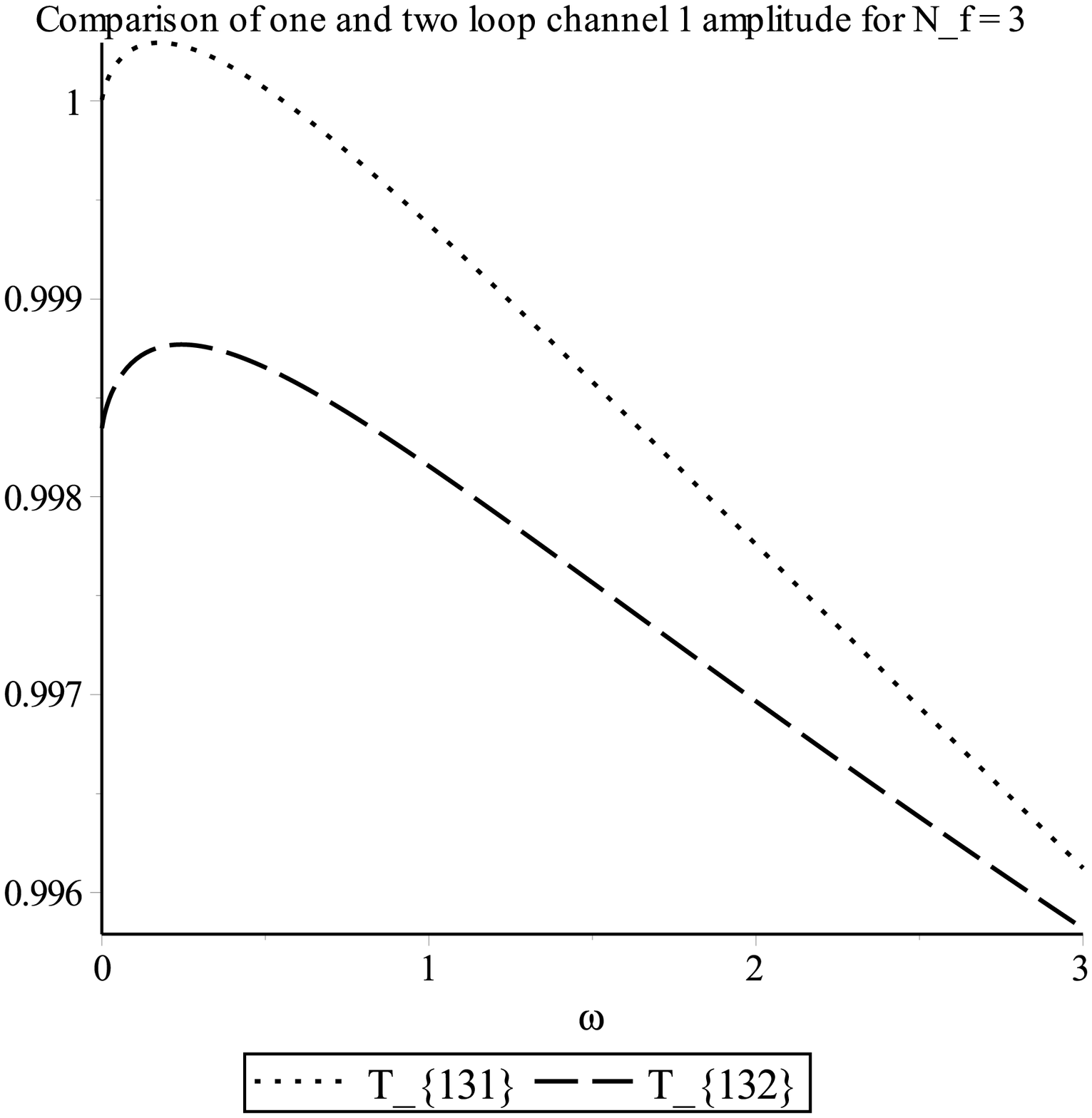}
~~~~
\includegraphics[width=7.6cm,height=7cm]{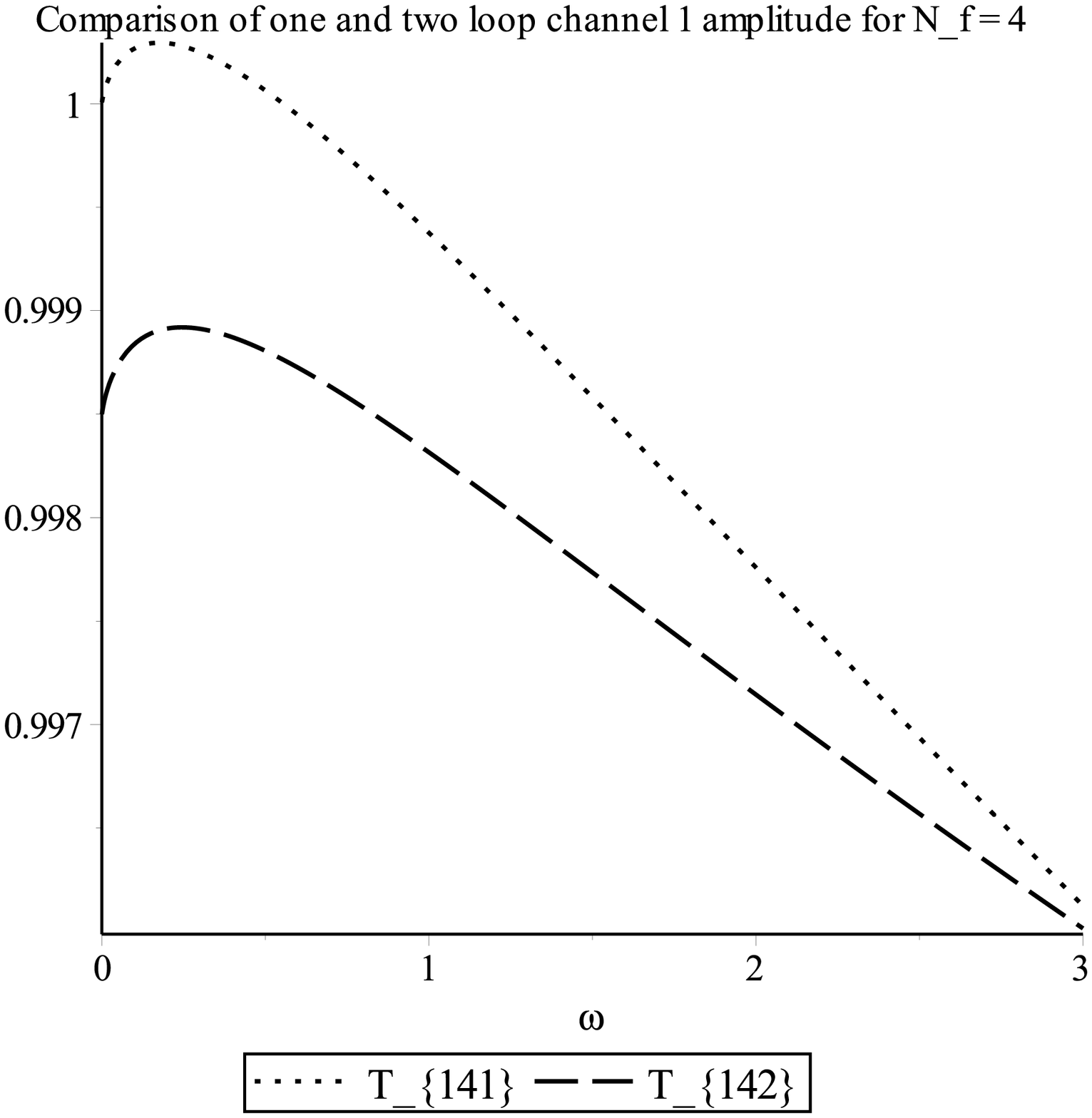}
\vspace{0.8cm}
\includegraphics[width=7.6cm,height=7cm]{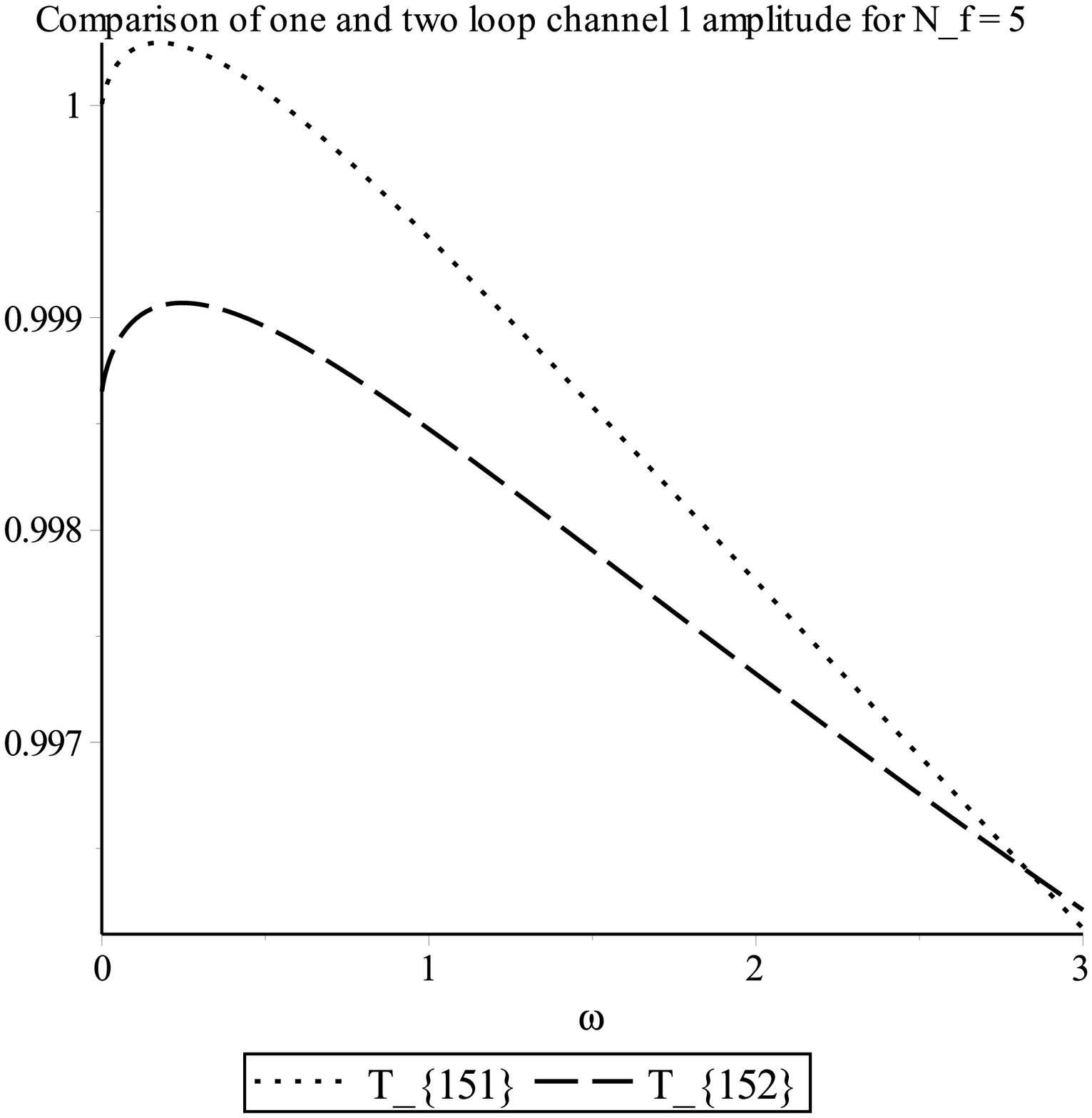}
\quad
\includegraphics[width=7.6cm,height=7cm]{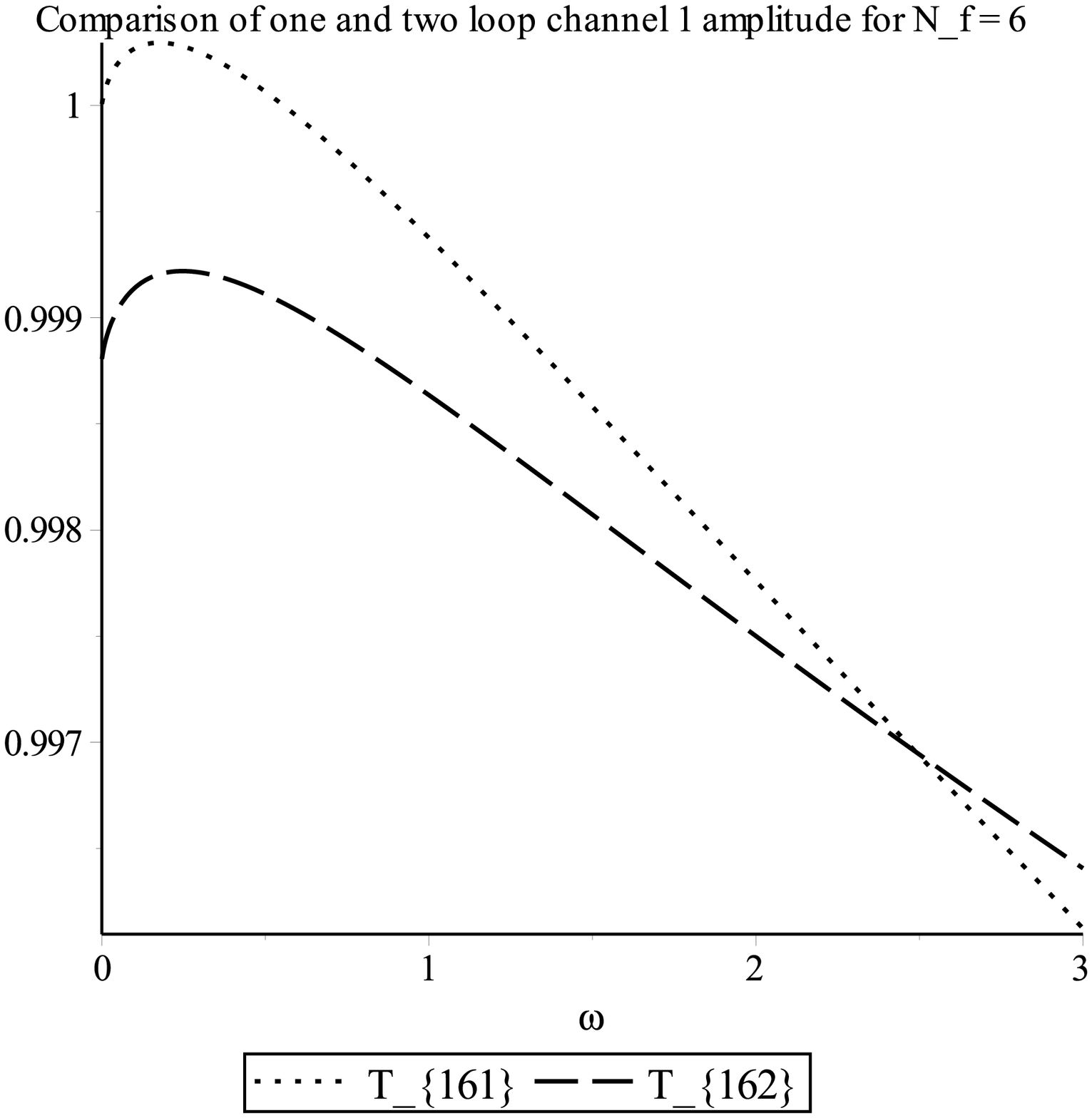}
\caption{Channel $1$ amplitude for tensor operator for different $\Nf$ values
and $\alpha_s$~$=$~$0.125$.}
\end{figure}}

\sect{Wilson operator.}

We devote this section to the results for the second moment of the Wilson
operator, $W_2$. This is because the renormalization of the two basic operators
with the same quantum numbers is not multiplicative unlike the operators of the
previous section. Instead $W_2$ and $\partial W_2$ mix under renormalization
and lead to a mixing matrix of anomalous dimensions which we will denote by
$\gamma_{ij}^{W_2}(a)$, \cite{8}. It is derived from the set of associated 
renormalization constants $Z_{ij}^{W_2}$ which are defined by 
\begin{equation}
{\cal O}_{\mbox{\footnotesize{o}}\, i} ~=~ Z^{W_2}_{ij} {\cal O}_j
\end{equation}
where the subscript ${}_{\mbox{\footnotesize{o}}}$ indicates a bare quantity.
For the two independent operators in the basis we have chosen it transpires
that $Z_{ij}^{W_2}$ is upper triangular. Specifically, \cite{8}, 
\begin{equation}
Z^{W_2}_{ij} ~=~ \left(
\begin{array}{cc}
Z^{W_2}_{11} & Z^{W_2}_{12} \\
0 & Z^{W_2}_{22} \\
\end{array}
\right) 
\end{equation}
which not only simplifies the renormalization process but allows us to make
contact with the usual way the renormalization of the operator $W_2$ is
derived. For instance, $W_2$ is ordinarily regarded as being a multiplicatively
renormalizable operator and $\gamma_{11}^{W_2}(a)$ is known to three loops in
the $\MSbar$ scheme, \cite{1,2,3}. However, this multiplicative renormalization
is due to the fact that when one renormalizes the operator it is invariably in 
the situation where it is inserted in a quark $2$-point function at zero 
momentum. This is not the momentum configuration we use here. Therefore, in the
zero momentum operator insertion setup the mixing with the total derivative 
operator is not accessible. This is because momentum conservation would render 
the Feynman rule for $\partial W_2$ to be zero. Another way of putting this is 
that the full content of the renormalization of the $W_2$ sector can only be 
fully appreciated at a non-exceptional point. An alternative basis to the one 
chosen here, such as that with the flavour non-singlet operator 
${\cal S} \left( D^\mu \bar{\psi} \right) \gamma^\nu \psi$ instead, would lead 
to a different form of the mixing matrix. So proceeding with our basis choice,
\cite{8} and following the projection method given earlier we have constructed 
all the amplitudes for both operators $W_2$ and $\partial W_2$ to two loops in 
the $\MSbar$ scheme. As the first stage we have verified that the correct two 
loop $\MSbar$ renormalization constants emerge in both cases. For the former 
operator the tally is with \cite{1,2}. For $\partial W_2$ we have verified 
that, \cite{8}, 
\begin{equation}
\left. \frac{}{} \gamma^{W_2}_{22}(a) \right|_{\MSbars} ~=~ O(a^3) 
\end{equation}
which is the same as $\gamma^V(a)$. This is not unconnected as $\partial W_2$
is the total derivative of the non-singlet vector current. So it is reassuring
that this emerges consistently. Partly related to this is that the amplitudes 
for this operator obey relations similar to (\ref{Vrel}). In particular we have
checked to two loops that, \cite{8},  
\begin{eqnarray}
\left. \Sigma^{\partial W_2}_{(1)}(p,q) \right|_{\omega = 1} &=& 
\left. \Sigma^{\partial W_2}_{(2)}(p,q) \right|_{\omega = 1} ~~~,~~~
\left. \Sigma^{\partial W_2}_{(3)}(p,q) \right|_{\omega = 1} ~=~ 
\left. \Sigma^{\partial W_2}_{(8)}(p,q) \right|_{\omega = 1} \nonumber \\
\left. \Sigma^{\partial W_2}_{(4)}(p,q) \right|_{\omega = 1} &=&
\left. \Sigma^{\partial W_2}_{(7)}(p,q) \right|_{\omega = 1} ~~~,~~~ 
\left. \Sigma^{\partial W_2}_{(5)}(p,q) \right|_{\omega = 1} ~=~
\left. \Sigma^{\partial W_2}_{(6)}(p,q) \right|_{\omega = 1} \nonumber \\
\left. \Sigma^{\partial W_2}_{(9)}(p,q) \right|_{\omega = 1} &=& 
\left. \Sigma^{\partial W_2}_{(10)}(p,q) \right|_{\omega = 1} 
\end{eqnarray}
which are established from the left-right symmetry of the underlying Green's
function for arbitrary $\omega$. For the operator $W_2$ itself there are no
similar relations. This is because the covariant derivative in the operator
definition only acts on the quark and not the anti-quark field. 

While we have presented results for other operators graphically we will do so
for $W_2$ too. However, as a guide to the form of what lies behind such a
representation we provide the channel $2$ amplitude in analytic form in the
$\MSbar$ scheme. We have 
\begin{eqnarray}
\left. \Sigma^{W_2}_{(2)}(p,q) \right|_\omega^{\alpha=0} &=& -~ 1 \nonumber \\
&& + \left[
\frac{79}{18}
+ \frac{5}{3} \frac{1}{[\omega-4]}
- \frac{17}{6} \ln(\omega)
- 10 \ln(\omega) \frac{1}{[\omega-4]^2}
- \frac{28}{3} \ln(\omega) \frac{1}{[\omega-4]}
\right. \nonumber \\
&& \left. ~~~
+ \Phi_{(1)\,\omega,\omega}
- \frac{9}{8} \Phi_{(1)\,\omega,\omega} \frac{1}{\omega}
+ \frac{5}{2} \Phi_{(1)\,\omega,\omega} \frac{1}{[\omega-4]^2}
+ \frac{17}{8} \Phi_{(1)\,\omega,\omega} \frac{1}{[\omega-4]}
\right] C_F a
\nonumber \\
&& +~ \left[ 
\left[ 
- \frac{2261}{162}
- 6 \frac{1}{[\omega-4]}
+ \frac{92}{9} \ln(\omega)
+ 36 \ln(\omega) \frac{1}{[\omega-4]^2}
+ \frac{310}{9} \ln(\omega) \frac{1}{[\omega-4]}
\right. \right. \nonumber \\
&& \left. \left. ~~~~~~
- \frac{31}{9} \Phi_{(1)\,\omega,\omega}
+ \frac{139}{36} \Phi_{(1)\,\omega,\omega} \frac{1}{\omega}
- 9 \Phi_{(1)\,\omega,\omega} \frac{1}{[\omega-4]^2}
\right. \right. \nonumber \\
&& \left. \left. ~~~~~~
- \frac{283}{36} \Phi_{(1)\,\omega,\omega} \frac{1}{[\omega-4]}
\right] 
C_F T_F \Nf 
\right. \nonumber \\
&& \left. +~ \left[
\frac{24337}{648}
+ \frac{241}{18} \frac{1}{[\omega-4]}
+ 18 \zeta_3 \frac{1}{[\omega-4]^2}
+ 21 \zeta_3 \frac{1}{[\omega-4]}
+ \zeta_3
- \frac{2209}{72} \ln(\omega)
\right. \right. \nonumber \\
&& \left. \left. ~~~~~
- \frac{241}{3} \ln(\omega) \frac{1}{[\omega-4]^2}
- \frac{1141}{12} \ln(\omega) \frac{1}{[\omega-4]}
- \frac{29}{8} \ln^2(\omega)
\right. \right. \nonumber \\
&& \left. \left. ~~~~~
- \frac{143}{2} \ln^2(\omega) \frac{1}{[\omega-4]^2}
- \frac{61}{2} \ln^2(\omega) \frac{1}{[\omega-4]}
+ \frac{1}{4} \ln(\omega) \Phi_{(1)\,\omega,\omega}
\right. \right. \nonumber \\
&& \left. \left. ~~~~~
- \frac{15}{32} \ln(\omega) \Phi_{(1)\,\omega,\omega} \frac{1}{\omega}
+ \frac{143}{8} \ln(\omega) \Phi_{(1)\,\omega,\omega} \frac{1}{[\omega-4]^2}
\right. \right. \nonumber \\
&& \left. \left. ~~~~~
+ \frac{271}{32} \ln(\omega) \Phi_{(1)\,\omega,\omega} \frac{1}{[\omega-4]}
- \frac{31}{12} \Omega_{(2)\,\omega,\omega}
- \frac{25}{32} \Omega_{(2)\,\omega,\omega} \frac{1}{\omega}
\right. \right. \nonumber \\
&& \left. \left. ~~~~~
- \frac{41}{8} \Omega_{(2)\,\omega,\omega} \frac{1}{[\omega-4]^2}
- \frac{737}{96} \Omega_{(2)\,\omega,\omega} \frac{1}{[\omega-4]}
- \frac{17}{3} \Omega_{(2)\,1,\omega}
\right. \right. \nonumber \\
&& \left. \left. ~~~~~
- 59 \Omega_{(2)\,1,\omega} \frac{1}{[\omega-4]^2}
- \frac{205}{6} \Omega_{(2)\,1,\omega} \frac{1}{[\omega-4]}
+ \frac{551}{36} \Phi_{(1)\,\omega,\omega}
\right. \right. \nonumber \\
&& \left. \left. ~~~~~~
- \frac{1039}{72} \Phi_{(1)\,\omega,\omega} \frac{1}{\omega}
+ \frac{241}{12} \Phi_{(1)\,\omega,\omega} \frac{1}{[\omega-4]^2}
+ \frac{2683}{72} \Phi_{(1)\,\omega,\omega} \frac{1}{[\omega-4]}
\right. \right. \nonumber \\
&& \left. \left. ~~~~~~
+ \frac{1}{2} \Phi_{(1)\,\omega,\omega}^2
- \frac{1}{2} \Phi_{(1)\,\omega,\omega}^2 \frac{1}{\omega}
- 2 \Phi_{(2)\,\omega,\omega}
+ \frac{25}{16} \Phi_{(2)\,\omega,\omega} \frac{1}{\omega^2}
\right. \right. \nonumber \\
&& \left. \left. ~~~~~~
+ \frac{29}{16} \Phi_{(2)\,\omega,\omega} \frac{1}{\omega}
- \frac{29}{16} \Phi_{(2)\,\omega,\omega} \frac{1}{[\omega-4]^2}
- \frac{45}{16} \Phi_{(2)\,\omega,\omega} \frac{1}{[\omega-4]}
\right. \right. \nonumber \\
&& \left. \left. ~~~~~~
+ 26 \Phi_{(2)\,1,\omega}
+ 224 \Phi_{(2)\,1,\omega} \frac{1}{[\omega-4]^2}
+ 160 \Phi_{(2)\,1,\omega} \frac{1}{[\omega-4]}
\right. \right. \nonumber \\
&& \left. \left. ~~~~~~
+ 3 \Phi_{(2)\,1,\omega} \omega
\right] C_F C_A 
\right. \nonumber \\
&& \left. +~ \left[
- \frac{10861}{648}
- \frac{44}{9} \frac{1}{[\omega-4]}
- 120 \zeta_3 \frac{1}{[\omega-4]^2}
- 112 \zeta_3 \frac{1}{[\omega-4]}
- 34 \zeta_3
\right. \right. \nonumber \\
&& \left. \left. ~~~~~
+ \frac{221}{12} \ln(\omega)
+ \frac{88}{3} \ln(\omega) \frac{1}{[\omega-4]^2}
+ \frac{1001}{18} \ln(\omega) \frac{1}{[\omega-4]}
+ \frac{65}{18} \ln(\omega)^2
\right. \right. \nonumber \\
&& \left. \left. ~~~~~
+ \frac{454}{3} \ln^2(\omega) \frac{1}{[\omega-4]^2}
+ \frac{490}{9} \ln^2(\omega) \frac{1}{[\omega-4]}
+ \frac{4}{3} \ln(\omega) \Phi_{(1)\,\omega,\omega}
\right. \right. \nonumber \\
&& \left. \left. ~~~~~
- \frac{13}{24} \ln(\omega) \Phi_{(1)\,\omega,\omega} \frac{1}{\omega}
- \frac{227}{6} \ln(\omega) \Phi_{(1)\,\omega,\omega} \frac{1}{[\omega-4]^2}
\right. \right. \nonumber \\
&& \left. \left. ~~~~~
- \frac{379}{24} \ln(\omega) \Phi_{(1)\,\omega,\omega} \frac{1}{[\omega-4]}
+ \frac{8}{3} \Omega_{(2)\,\omega,\omega}
+ \frac{3}{4} \Omega_{(2)\,\omega,\omega} \frac{1}{\omega}
+ \Omega_{(2)\,\omega,\omega} \frac{1}{[\omega-4]^2}
\right. \right. \nonumber \\
&& \left. \left. ~~~~~
+ \frac{79}{12} \Omega_{(2)\,\omega,\omega} \frac{1}{[\omega-4]}
+ \frac{40}{3} \Omega_{(2)\,1,\omega}
+ 148 \Omega_{(2)\,1,\omega} \frac{1}{[\omega-4]^2}
\right. \right. \nonumber \\
&& \left. \left. ~~~~~
+ \frac{244}{3} \Omega_{(2)\,1,\omega} \frac{1}{[\omega-4]}
- \frac{679}{36} \Phi_{(1)\,\omega,\omega}
+ \frac{487}{72} \Phi_{(1)\,\omega,\omega} \frac{1}{\omega}
\right. \right. \nonumber \\
&& \left. \left. ~~~~~
- \frac{22}{3} \Phi_{(1)\,\omega,\omega} \frac{1}{[\omega-4]^2}
- \frac{4039}{72} \Phi_{(1)\,\omega,\omega} \frac{1}{[\omega-4]}
- \Phi_{(1)\,\omega,\omega}^2
+ \Phi_{(1)\,\omega,\omega}^2 \frac{1}{\omega}
\right. \right. \nonumber \\
&& \left. \left. ~~~~~
- \frac{3}{2} \Phi_{(2)\,\omega,\omega} \frac{1}{\omega^2}
+ \Phi_{(2)\,\omega,\omega} \frac{1}{\omega}
- \frac{9}{2} \Phi_{(2)\,\omega,\omega} \frac{1}{[\omega-4]^2}
- \Phi_{(2)\,\omega,\omega} \frac{1}{[\omega-4]}
\right. \right. \nonumber \\
&& \left. \left. ~~~~~~
- 52 \Phi_{(2)\,1,\omega}
- 512 \Phi_{(2)\,1,\omega} \frac{1}{[\omega-4]^2}
- 336 \Phi_{(2)\,1,\omega} \frac{1}{[\omega-4]}
\right. \right. \nonumber \\
&& \left. \left. ~~~~~
- 4 \Phi_{(2)\,1,\omega} \omega
\right] C_F^2 
\right] a^2 ~+~ O(a^3) ~. 
\label{w2amp2}
\end{eqnarray}
The reason for concentrating on channel $2$ is that that is the channel which
corresponds to the Feynman rule for $W_2$. By contrast the channel $1$ basis
tensor involves the external momentum $p$ which would be associated with the
covariant derivative in the operator ${\cal S} \left( D^\mu \bar{\psi} \right) 
\gamma^\nu \psi$ which is not in our basis. Graphically we have presented the
results for channel $2$ in Figure $4$. Although strictly we have plotted the
negative of (\ref{w2amp2}) in order to easily compare with the Figures of the
previous section. In essence they are formally similar to earlier 
representations. For instance, as $\Nf$ increases there is less variation 
between the one and two loop amplitudes as a function of $\omega$ for the value
of $\alpha_s$ considered. Finally, we have checked that the analogous plots for
$\partial W_2$ are the same as those for the vector case of the previous 
section. This is not unexpected given the structural similarity of the two 
operators. Therefore, we have not produced parallel graphs for $\partial W_2$. 

{\begin{figure}
\includegraphics[width=7.6cm,height=7cm]{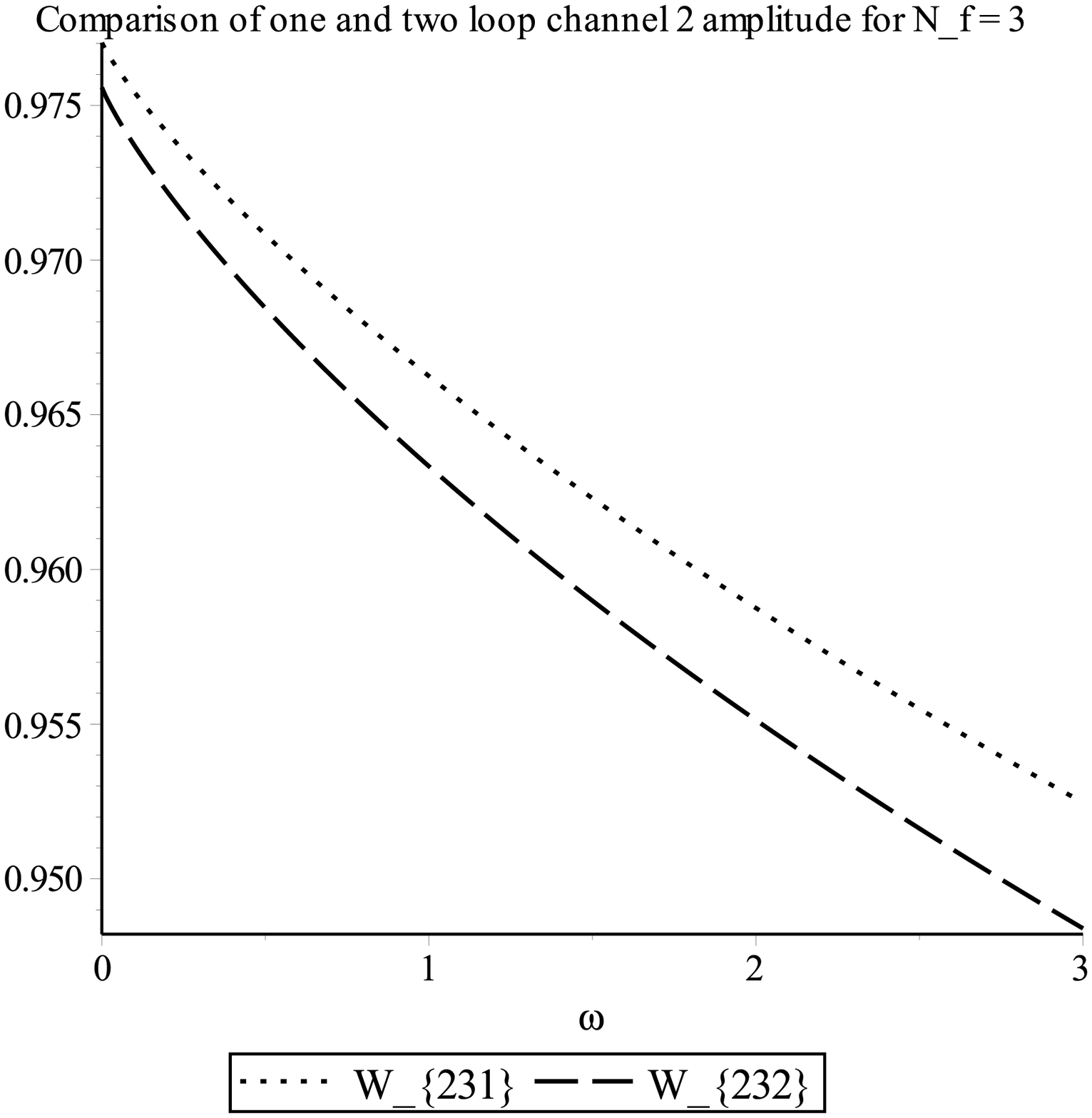}
~~~~
\includegraphics[width=7.6cm,height=7cm]{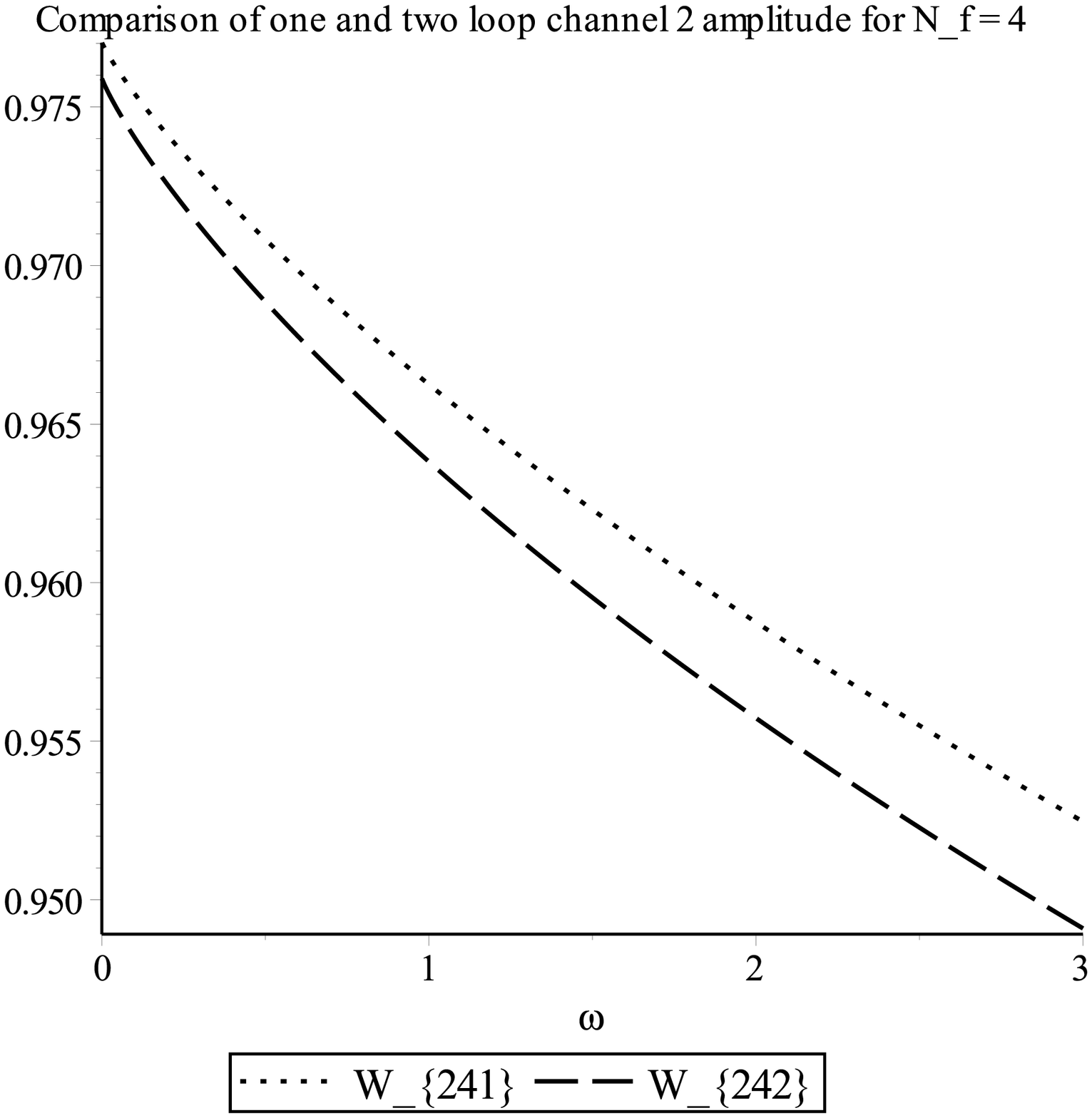}
\vspace{0.8cm}
\includegraphics[width=7.6cm,height=7cm]{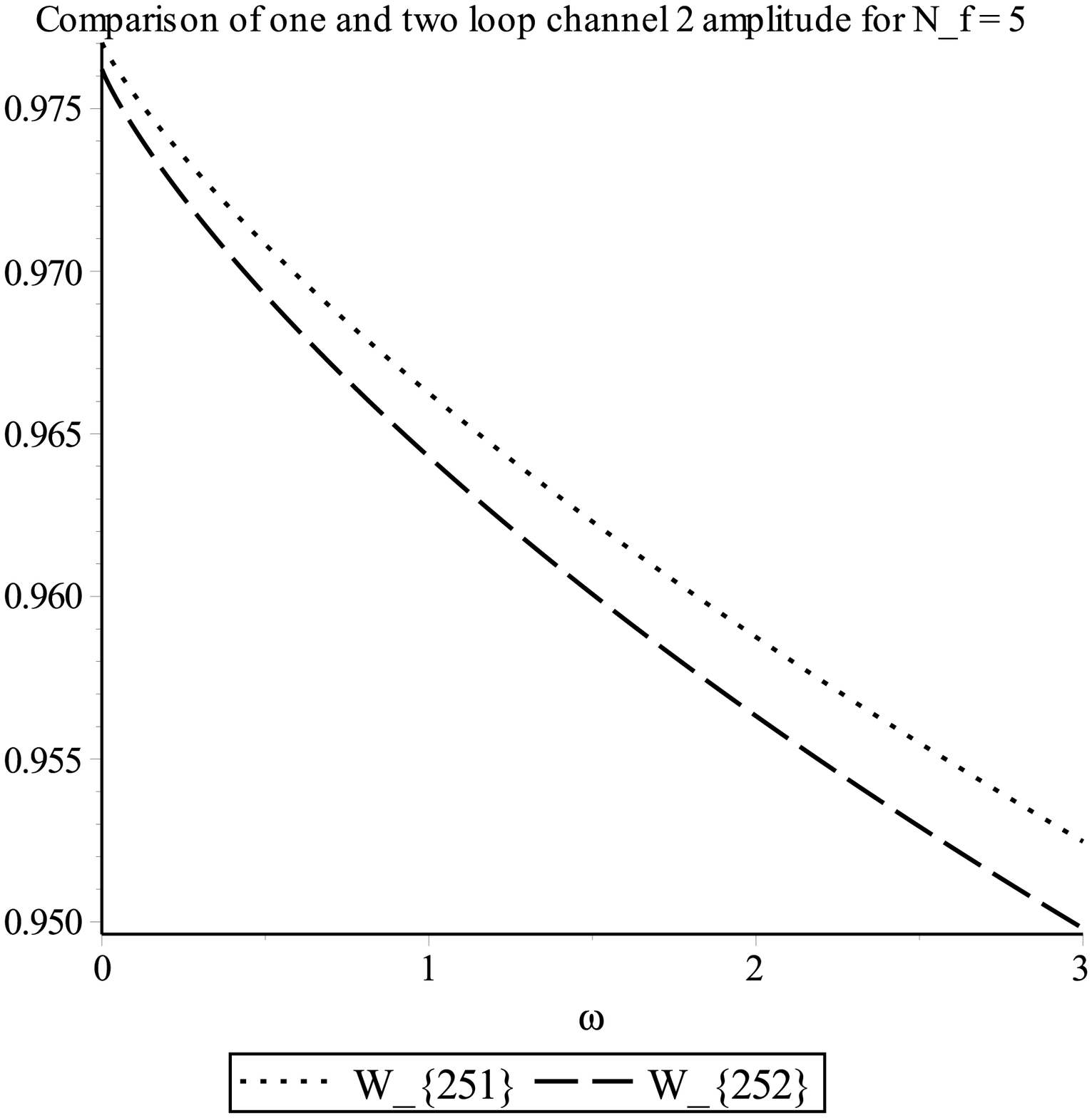}
\quad
\includegraphics[width=7.6cm,height=7cm]{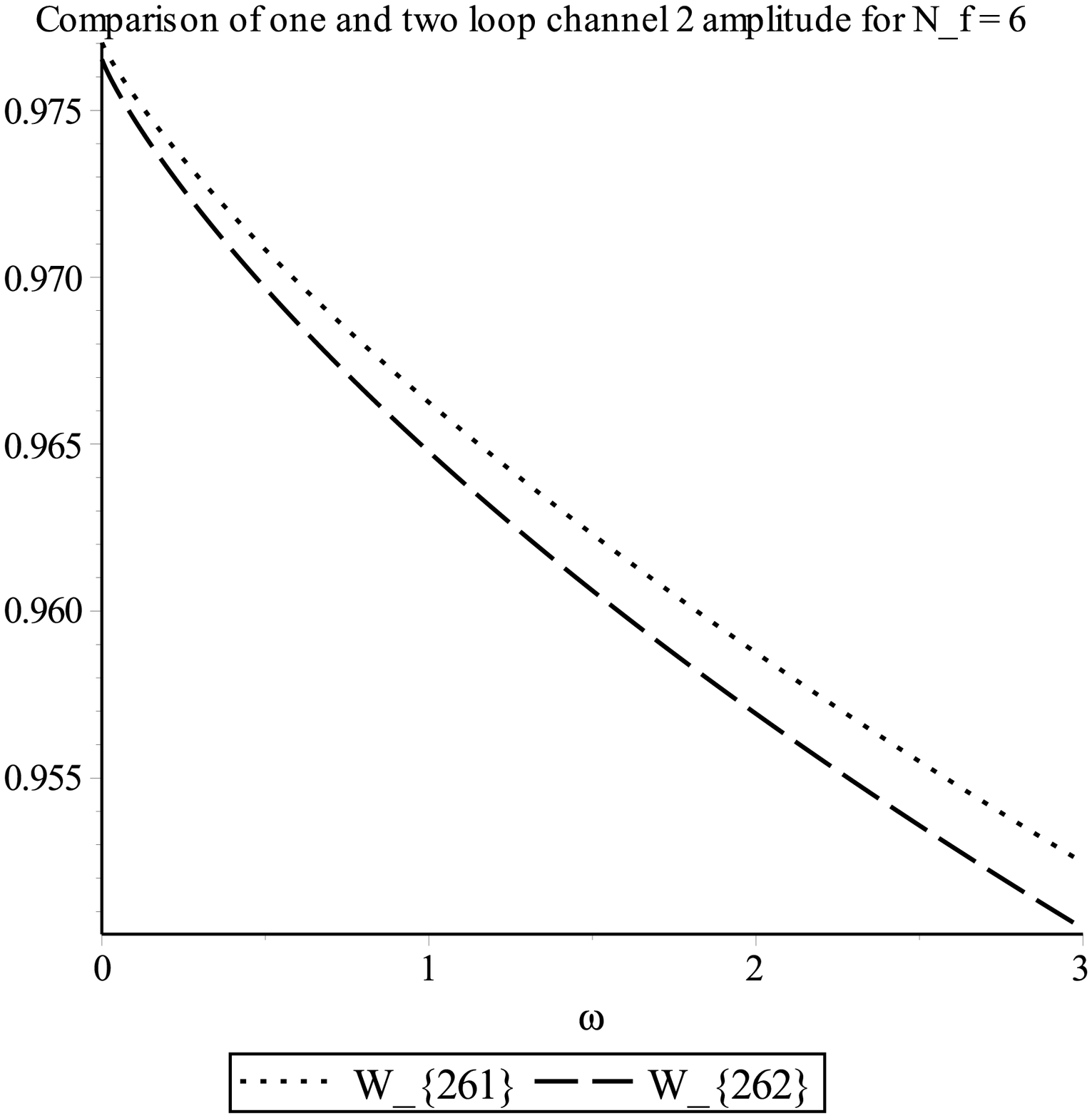}
\caption{Channel $2$ amplitude for $W_2$ operator for different $\Nf$ values
and $\alpha_s$~$=$~$0.125$.}
\end{figure}}

\sect{Discussion.}

We finish with brief observations. The inclusion of a parameter, $\omega$,
which allows one to tune the momentum flow through the operator inserted in the
basic Green's function should prove useful for lattice analyses. For instance,
it provides a tool with which one can reduce the errors on measurements of 
structure functions. Although from the various graphs the variation on the
highlighted amplitudes over the range of $\omega$ plotted is not the same
overall. For instance, for the vector and tensor operators the two loop
correction is virtually insignificant compared to the scalar case. For the 
vector this is important in the context of $W_2$ itself because the behaviour
of the vector current amplitudes are effectively the same as those for
$\partial W_2$. The mixing between these operators has to be disentangled to
obtain a clear signal for the basic operator underlying the nucleon structure
function measurements. That the variation between one and two loops is small
suggests that an error estimate could be commensurate. The present computations
complete the two loop analysis of operator renormalization which was begun in 
\cite{4,5,6,7,8}. The next stage in any programme of this nature would be to
go to the next loop order in order to improve the precision even more. This is
a major step as one requires the underlying three loop master integrals.  

\vspace{1cm}
\noindent
{\bf Acknowledgements.} This work was carried out with the support in part from
the STFC Consolidated Grant ST/L000431/1 and STFC studentship (JMB) as well as 
through a John Lennon Memorial Scholarship (JMB). The authors thanks R. 
Horsley, P.E.L. Rakow and G. Schierholz for useful discussions.

\appendix

\sect{Bases and projection matrices.}

In this appendix we record the full forms of the basis tensors for the Green's
function with the operator insertion as well as the projection matrices. As we
are using quark operators we have to be aware that the basis tensors do not
only depend on $\eta_{\mu\nu}$ and the two independent external momenta $p$ and
$q$ but also products of $\gamma$-matrices. The latter can also carry free 
Lorentz indices but are necessary in order to carry the spinor indices. In 
order to achieve this we use the generalized $d$-dimensional $\gamma$-matrices 
defined by, \cite{47,48,49,50,51}, 
\begin{equation}
\Gamma_{(n)}^{\mu_1 \ldots \mu_n} ~=~ \gamma^{[\mu_1} \ldots \gamma^{\mu_n]} ~.
\end{equation}
Our convention is that the $1/n!$ factor is included in the antisymmetrization.
As these objects are totally antisymmetric in the Lorentz indices then the
tensor basis for various operators have natural partitions which is manifested
in results such as
\begin{equation}
\mbox{tr} \left( \Gamma_{(m)}^{\mu_1 \ldots \mu_m}
\Gamma_{(n)}^{\nu_1 \ldots \nu_n} \right) ~ \propto ~ \delta_{mn}
I^{\mu_1 \ldots \mu_m \nu_1 \ldots \nu_n} 
\end{equation}
where $I^{\mu_1 \ldots \mu_m \nu_1 \ldots \nu_n}$ is the generalized unit
matrix, \cite{49,50,51}. Equipped with these structures the basic tensors are 
in essence the same as those for the symmetric point but the projection 
matrices are $\omega$ dependent. In the $\omega$~$\rightarrow$~$1$ limit we 
have checked that the matrices of \cite{7,8} are reproduced. For the scalar, 
vector and tensor operators the bases are, \cite{7}, 
\begin{equation}
{\cal P}^{S}_{(1)}(p,q) ~=~ \Gamma_{(0)} ~~~,~~~
{\cal P}^{S}_{(2)}(p,q) ~=~ \frac{1}{\mu^2} \Gamma_{(2)}^{pq}
\end{equation}
\begin{eqnarray}
{\cal P}^{V}_{(1) \mu }(p,q) &=& \gamma_\mu ~~~,~~~
{\cal P}^{V}_{(2) \mu }(p,q) ~=~ \frac{{p}^\mu \pslash}{\mu^2} ~~~,~~~
{\cal P}^{V}_{(3) \mu }(p,q) ~=~ \frac{{p}_\mu \qslash}{\mu^2} ~, \nonumber \\
{\cal P}^{V}_{(4) \mu }(p,q) &=& \frac{{q}_\mu \pslash}{\mu^2} ~~~,~~~
{\cal P}^{V}_{(5) \mu }(p,q) ~=~ \frac{{q}_\mu \qslash}{\mu^2} ~~~,~~~
{\cal P}^{V}_{(6) \mu }(p,q) ~=~ \frac{1}{\mu^2} \Gamma_{(3) \, \mu p q} 
\end{eqnarray}
and
\begin{eqnarray}
{\cal P}^{T}_{(1) \mu \nu }(p,q) &=& \Gamma_{(2) \, \mu\nu} ~~~,~~~
{\cal P}^{T}_{(2) \mu \nu }(p,q) ~=~ \frac{1}{\mu^2} \left[ p_\mu q_\nu - p_\nu
q_\mu \right] \Gamma_{(0)} ~, \nonumber \\
{\cal P}^{T}_{(3) \mu \nu }(p,q) &=& \frac{1}{\mu^2}
\left[ \Gamma_{(2) \, \mu p} p_\nu - \Gamma_{(2) \, \nu p} p_\mu
\right] ~~~,~~~
{\cal P}^{T}_{(4) \mu \nu }(p,q) ~=~ \frac{1}{\mu^2}
\left[ \Gamma_{(2) \, \mu p} q_\nu - \Gamma_{(2) \, \nu p} q_\mu
\right] ~, \nonumber \\
{\cal P}^{T}_{(5) \mu \nu }(p,q) &=& \frac{1}{\mu^2}
\left[ \Gamma_{(2) \, \mu q} p_\nu - \Gamma_{(2) \, \nu q} p_\mu
\right] ~~~,~~~
{\cal P}^{T}_{(6) \mu \nu }(p,q) ~=~ \frac{1}{\mu^2}
\left[ \Gamma_{(2) \, \mu q} q_\nu - \Gamma_{(2) \, \nu q} q_\mu
\right] ~, \nonumber \\
{\cal P}^{T}_{(7) \mu \nu }(p,q) &=& \frac{1}{\mu^4}
\left[ \Gamma_{(2) \, p q} p_\mu q_\nu - \Gamma_{(2) \, p q} p_\nu q_\mu
\right] ~~~,~~~
{\cal P}^{T}_{(8) \mu \nu }(p,q) ~=~ \frac{1}{\mu^2}
\Gamma_{(4) \, \mu \nu p q} 
\end{eqnarray}
respectively where the contraction of a $\Gamma$-matrix with an external
momentum is represented by the momentum itself replacing the contracting index. 
The situation for the Wilson operator is slightly different. As in \cite{8} we 
choose a set of tensors which have the same symmetry properties as the
operators themselves. However, in order to ensure tracelessness the associated 
contractions in certain tensors involves $pq$ and hence $\omega$. So unlike 
$S$, $V$ and $T$ certain tensors for the $W_2$ and $\partial W_2$
decomposition are $\omega$ dependent. In particular
\begin{eqnarray}
{\cal P}^{W_2}_{(1) \mu \nu }(p,q) &=& \gamma_\mu p_\nu + \gamma_\nu p_\mu
- \frac{2}{d} \pslash \eta_{\mu\nu} ~~,~~
{\cal P}^{W_2}_{(2) \mu \nu }(p,q) ~=~ \gamma_\mu q_\nu + \gamma_\nu q_\mu
- \frac{2}{d} \qslash \eta_{\mu\nu} \nonumber \\
{\cal P}^{W_2}_{(3) \mu \nu }(p,q) &=& \pslash \left[
\frac{1}{\mu^2} p_\mu p_\nu + \frac{1}{d} \eta_{\mu\nu} \right] ~~,~~
{\cal P}^{W_2}_{(4) \mu \nu }(p,q) ~=~ \pslash \left[
\frac{1}{\mu^2} p_\mu q_\nu + \frac{1}{\mu^2} q_\mu p_\nu
- \frac{(2-\omega)}{d} \eta_{\mu\nu} \right] \nonumber \\
{\cal P}^{W_2}_{(5) \mu \nu }(p,q) &=& \pslash \left[
\frac{1}{\mu^2} q_\mu q_\nu + \frac{1}{d} \eta_{\mu\nu} \right] ~~,~~
{\cal P}^{W_2}_{(6) \mu \nu }(p,q) ~=~ \qslash \left[
\frac{1}{\mu^2} p_\mu p_\nu + \frac{1}{d} \eta_{\mu\nu} \right] \nonumber \\
{\cal P}^{W_2}_{(7) \mu \nu }(p,q) &=& \qslash \left[
\frac{1}{\mu^2} p_\mu q_\nu + \frac{1}{\mu^2} q_\mu p_\nu
-\frac{(2-\omega)}{d} \eta_{\mu\nu} \right] ~,~
{\cal P}^{W_2}_{(8) \mu \nu }(p,q) ~=~ \qslash \left[
\frac{1}{\mu^2} q_\mu q_\nu + \frac{1}{d} \eta_{\mu\nu} \right] \nonumber \\
{\cal P}^{W_2}_{(9) \mu \nu }(p,q) &=& \frac{1}{\mu^2} \left[
\Gamma_{(3) \, \mu p q } p_\nu + \Gamma_{(3) \, \nu p q } p_\mu \right]
\nonumber \\
{\cal P}^{W_2}_{(10) \mu \nu }(p,q) &=& \frac{1}{\mu^2} \left[
\Gamma_{(3) \, \mu p q } q_\nu + \Gamma_{(3) \, \nu p q } q_\mu \right] ~.
\end{eqnarray}
We note that in the renormalization of the operator $W_2$ the basis element
corresponding to the tree term of (\ref{gfun}) is channel $2$ and not $1$. This
is because the covariant derivative acts on the quark field and not the
anti-quark. For $\partial W_2$ both channels $1$ and $2$ are relevant to the
operator renormalization.  

Equipped with these the various projection matrices are 
\begin{equation}
{\cal M}^S ~=~ \frac{1}{4\omega[\omega-4]} \left(
\begin{array}{cc}
\omega [\omega-4] & 0 \\
0 & 4 \\
\end{array}
\right) 
\end{equation}
for the scalar case. For the remaining operators since the projection matrices 
are symmetric by construction we list the upper triangular elements only. For 
the vector operator if we factor off a common factor, which is not the 
determinant of ${\cal N}^V$, via
\begin{equation}
{\cal M}^V ~=~ \frac{1}{4(d-2)\omega^2[\omega-4]^2} 
\tilde{\cal M}^V 
\end{equation}
then the elements are
\begin{eqnarray}
\tilde{\cal M}^V_{11} &=& [\omega - 4]^2 \omega^2 ~~,~~
\tilde{\cal M}^V_{12} ~=~ - 4 [\omega - 4] \omega ~~,~~
\tilde{\cal M}^V_{13} ~=~ 2 [\omega - 2] [\omega - 4] \omega \nonumber \\
\tilde{\cal M}^V_{14} &=& 2 [\omega - 2] [\omega - 4] \omega ~~,~~
\tilde{\cal M}^V_{15} ~=~ - 4 [\omega - 4] \omega ~~,~~
\tilde{\cal M}^V_{16} ~=~ 0 ~~,~~
\tilde{\cal M}^V_{22} ~=~ 16 [d - 1] \nonumber \\
\tilde{\cal M}^V_{23} &=& - 8 [d - 1] [\omega - 2] ~~,~~
\tilde{\cal M}^V_{24} ~=~ - 8 [d - 1] [\omega - 2] \nonumber \\
\tilde{\cal M}^V_{25} &=& - 4 [2 [\omega^2 - 4 \omega + 2] 
- [\omega - 2]^2 d] ~~,~~
\tilde{\cal M}^V_{26} ~=~ 0 ~~,~~
\tilde{\cal M}^V_{33} ~=~ 4 [\omega^2 - 4 \omega - 4 + 4 d] \nonumber \\
\tilde{\cal M}^V_{34} &=& 4 [d - 1] [\omega - 2]^2 ~~,~~
\tilde{\cal M}^V_{35} ~=~ - 8 [d - 1] [\omega - 2] ~~,~~
\tilde{\cal M}^V_{36} ~=~ 0 \nonumber \\
\tilde{\cal M}^V_{44} &=& 4 [\omega^2 - 4 \omega - 4 + 4 d] ~~,~~
\tilde{\cal M}^V_{45} ~=~ - 8 [d - 1] [\omega - 2] ~~,~~
\tilde{\cal M}^V_{46} ~=~ 0 \nonumber \\
\tilde{\cal M}^V_{55} &=& 16 [d - 1] ~~,~~
\tilde{\cal M}^V_{56} ~=~ 0 ~~,~~
\tilde{\cal M}^V_{66} ~=~ 4 [\omega - 4] \omega ~.
\end{eqnarray}
For such a simple spin-$1$ operator the $\omega$ dependence is quite involved.
In the tensor case, with 
\begin{equation}
{\cal M}^T ~=~ \frac{1}{4(d-2)(d-3)\omega^2[\omega-4]^2} 
\tilde{\cal M}^T 
\end{equation}
then
\begin{eqnarray}
\tilde{\cal M}^T_{11} &=& - [\omega - 4]^2 \omega^2 ~~,~~
\tilde{\cal M}^T_{12} ~=~ 0 ~~,~~
\tilde{\cal M}^T_{13} ~=~ 4 [\omega - 4] \omega ~~,~~
\tilde{\cal M}^T_{14} ~=~ - 2 [\omega - 2] [\omega - 4] \omega \nonumber \\
\tilde{\cal M}^T_{15} &=& - 2 [\omega - 2] [\omega - 4] \omega ~~,~~
\tilde{\cal M}^T_{16} ~=~ 4 [\omega - 4] \omega ~~,~~
\tilde{\cal M}^T_{17} ~=~ 4 [\omega - 4] \omega ~~,~~
\tilde{\cal M}^T_{18} ~=~ 0 \nonumber \\
\tilde{\cal M}^T_{22} &=& - 2 [d - 2] [d - 3] [\omega - 4] \omega ~~,~~
\tilde{\cal M}^T_{23} ~=~ 0 ~~,~~
\tilde{\cal M}^T_{24} ~=~ 0 ~~,~~
\tilde{\cal M}^T_{25} ~=~ 0 ~~,~~
\tilde{\cal M}^T_{26} ~=~ 0 \nonumber \\
\tilde{\cal M}^T_{27} &=& 0 ~~,~~
\tilde{\cal M}^T_{28} ~=~ 0 ~~,~~
\tilde{\cal M}^T_{33} ~=~ - 8 [d - 1] ~~,~~
\tilde{\cal M}^T_{34} ~=~ 4 [d - 1] [\omega - 2] \nonumber \\
\tilde{\cal M}^T_{35} &=& 4 [d - 1] [\omega - 2] ~~,~~
\tilde{\cal M}^T_{36} ~=~ - 2 [d \omega^2 - 4 d \omega + 4 d - 3 \omega^2 
+ 12 \omega - 4] \nonumber \\
\tilde{\cal M}^T_{37} &=& - 8 [d - 1] ~~,~~
\tilde{\cal M}^T_{38} ~=~ 0 ~~,~~
\tilde{\cal M}^T_{44} ~=~ - 4 [2 d + \omega^2 - 4 \omega - 2] \nonumber \\
\tilde{\cal M}^T_{45} &=& - 2 [d - 1] [\omega - 2]^2 ~~,~~
\tilde{\cal M}^T_{46} ~=~ 4 [d - 1] [\omega - 2] ~~,~~
\tilde{\cal M}^T_{47} ~=~ 4 [d - 1] [\omega - 2] \nonumber \\
\tilde{\cal M}^T_{48} &=& 0 ~~,~~
\tilde{\cal M}^T_{55} ~=~ - 4 [2 d + \omega^2 - 4 \omega - 2] ~~,~~
\tilde{\cal M}^T_{56} ~=~ 4 [d - 1] [\omega - 2] \nonumber \\
\tilde{\cal M}^T_{57} &=& 4 [d - 1] [\omega - 2] ~~,~~
\tilde{\cal M}^T_{58} ~=~ 0 ~~,~~
\tilde{\cal M}^T_{66} ~=~ - 8 [d - 1] ~~,~~
\tilde{\cal M}^T_{67} ~=~ - 8 [d - 1] \nonumber \\
\tilde{\cal M}^T_{68} &=& 0 ~~,~~
\tilde{\cal M}^T_{77} ~=~ - 8 [d - 1] [d - 2] ~~,~~
\tilde{\cal M}^T_{78} ~=~ 0 ~~,~~
\tilde{\cal M}^T_{88} ~=~ - 4 [\omega - 4] \omega ~.
\end{eqnarray}
Finally, for the twist-$2$ operator we set
\begin{equation}
{\cal M}^{W_2} ~=~ \frac{1}{4(d-2)\omega^3[\omega-4]^3} 
\tilde{\cal M}^{W_2} 
\end{equation}
to produce the elements 
\begin{eqnarray}
\tilde{\cal M}^{W_2}_{11} &=& 2  [\omega - 4]^2 \omega^2 ~~,~~
\tilde{\cal M}^{W_2}_{12} ~=~ -  [\omega - 2] [\omega - 4]^2 \omega^2 ~~,~~
\tilde{\cal M}^{W_2}_{13} ~=~ - 16  [\omega - 4] \omega 
\nonumber \\
\tilde{\cal M}^{W_2}_{14} &=& 8  [\omega - 2] [\omega - 4] \omega ~~,~~
\tilde{\cal M}^{W_2}_{15} ~=~ - 4  [\omega - 2]^2 [\omega - 4] 
\omega ~~,~~ 
\tilde{\cal M}^{W_2}_{16} ~=~ 8  [\omega - 2] [\omega - 4] \omega
\nonumber \\
\tilde{\cal M}^{W_2}_{17} &=& - 2  [\omega^2 - 4 \omega + 8] 
[\omega - 4] \omega ~~,~~ 
\tilde{\cal M}^{W_2}_{18} ~=~ 8  [\omega - 2] [\omega - 4] \omega ~~,~~
\tilde{\cal M}^{W_2}_{19} ~=~ 0 \nonumber \\
\tilde{\cal M}^{W_2}_{110} &=& 0 ~~,~~ 
\tilde{\cal M}^{W_2}_{22} ~=~ 2  [\omega - 4]^2 \omega^2 ~~,~~
\tilde{\cal M}^{W_2}_{23} ~=~ 8  [\omega - 2] [\omega - 4] \omega
\nonumber \\
\tilde{\cal M}^{W_2}_{24} &=& - 2  [\omega^2 - 4 \omega + 8] 
[\omega - 4] \omega ~~,~~
\tilde{\cal M}^{W_2}_{25} ~=~ 8  [\omega - 2] [\omega - 4] \omega 
\nonumber \\
\tilde{\cal M}^{W_2}_{26} &=& - 4  [\omega - 2]^2 [\omega - 4] \omega ~~,~~
\tilde{\cal M}^{W_2}_{27} ~=~ 8  [\omega - 2] [\omega - 4] \omega ~~,~~
\tilde{\cal M}^{W_2}_{28} ~=~ - 16  [\omega - 4] \omega 
\nonumber \\
\tilde{\cal M}^{W_2}_{29} &=& 0 ~~,~~
\tilde{\cal M}^{W_2}_{210} ~=~ 0 ~~,~~ 
\tilde{\cal M}^{W_2}_{33} ~=~ 64 [d + 1] ~~,~~ 
\tilde{\cal M}^{W_2}_{34} ~=~ - 32 [d + 1]  [\omega - 2] 
\nonumber \\
\tilde{\cal M}^{W_2}_{35} &=& 16 [d \omega^2 - 4 d \omega + 4 d + 4] ~~,~~
\tilde{\cal M}^{W_2}_{36} ~=~ - 32 [d + 1]  [\omega - 2] \nonumber \\
\tilde{\cal M}^{W_2}_{37} &=& 16 [d \omega^2 - 4 d \omega + 4 d + 4] ~~,~~
\tilde{\cal M}^{W_2}_{38} ~=~ - 8 [d \omega^2 - 4 d \omega + 4 d - 2 \omega^2 
+ 8 \omega + 4]  [\omega - 2] 
\nonumber \\
\tilde{\cal M}^{W_2}_{39} &=& 0 ~~,~~
\tilde{\cal M}^{W_2}_{310} ~=~ 0 ~~,~~ 
\tilde{\cal M}^{W_2}_{44} ~-~ 8 [d \omega^2 - 4 d \omega + 8 d + 3 \omega^2 
- 12 \omega + 8]  \nonumber \\
\tilde{\cal M}^{W_2}_{45} &=& - 8 [4 d + \omega^2 - 4 \omega + 4]  
[\omega - 2] ~~,~~
\tilde{\cal M}^{W_2}_{46} ~=~ 16 [d + 1]  [\omega - 2]^2 \nonumber \\
\tilde{\cal M}^{W_2}_{47} &=& - 4 [d + 1]  [\omega^2 - 4 \omega + 8] 
[\omega - 2] ~~,~~
\tilde{\cal M}^{W_2}_{48} ~=~ 16 [d \omega^2 - 4 d \omega + 4 d + 4] 
 \nonumber \\
\tilde{\cal M}^{W_2}_{49} &=& 0 ~~,~~ 
\tilde{\cal M}^{W_2}_{410} ~=~ 0 ~~,~~
\tilde{\cal M}^{W_2}_{55} ~=~ 32 [2 d + \omega^2 - 4 \omega + 2] 
\nonumber \\
\tilde{\cal M}^{W_2}_{56} &=& - 8 [d \omega^2 - 4 d \omega + 4 d + 4]  
[\omega - 2] ~~,~~
\tilde{\cal M}^{W_2}_{57} ~=~ 16 [d + 1]  [\omega - 2]^2 \nonumber \\
\tilde{\cal M}^{W_2}_{58} &=& - 32 [d + 1]  [\omega - 2] ~~,~~
\tilde{\cal M}^{W_2}_{59} ~=~ 0 ~~,~~
\tilde{\cal M}^{W_2}_{510} ~=~ 0 ~~,~~ 
\tilde{\cal M}^{W_2}_{66} ~=~ 32 [2 d + \omega^2 - 4 \omega + 2]  
\nonumber \\
\tilde{\cal M}^{W_2}_{67} &=& - 8 [4 d + \omega^2 - 4 \omega + 4]  
[\omega - 2] ~~,~~
\tilde{\cal M}^{W_2}_{68} ~=~ 16 [d \omega^2 - 4 d \omega + 4 d + 4] ~~,~~
\tilde{\cal M}^{W_2}_{69} ~=~ 0 
\nonumber \\
\tilde{\cal M}^{W_2}_{610} &=& 0 ~~,~~ 
\tilde{\cal M}^{W_2}_{77} ~=~ 8 [d \omega^2 - 4 d \omega + 8 d + 3 \omega^2 - 12 \omega + 8]  ~~,~~
\tilde{\cal M}^{W_2}_{78} ~=~ - 32 [d + 1]  [\omega - 2] \nonumber \\
\tilde{\cal M}^{W_2}_{79} &=& 0 ~~,~~
\tilde{\cal M}^{W_2}_{710} ~=~ 0 ~~,~~
\tilde{\cal M}^{W_2}_{88} ~=~ 64 [d + 1]  ~~,~~
\tilde{\cal M}^{W_2}_{89} ~=~ 0 ~~,~~
\tilde{\cal M}^{W_2}_{810} ~=~ 0 \nonumber \\
\tilde{\cal M}^{W_2}_{99} &=& 8  [\omega - 4] \omega ~~,~~
\tilde{\cal M}^{W_2}_{910} ~=~ - 4  [\omega - 2] [\omega - 4] \omega ~~,~~ 
\tilde{\cal M}^{W_2}_{1010} ~=~ 8  [\omega - 4] \omega ~. 
\end{eqnarray}
For each case the $\omega$~$\rightarrow$~$0$ limit of the projection matrices
are singular as expected since this is a point where there are infrared
singularities.

\end{document}